\documentclass[acmsmall,screen,nonacm]{acmart}

\usepackage{soul}
\usepackage{comment}
\usepackage{pdflscape}

\usepackage{tcolorbox}
\tcbuselibrary{theorems}

\usepackage{tikz}
\usetikzlibrary{positioning,calc,fit} 

\usepackage{xcolor}
\usepackage{pgfplots}
\pgfplotsset{compat=1.17}
\usepackage{array}

\usepackage{adjustbox}   
\usepackage{booktabs} 
\usepackage{hyperref}

\newtcbtheorem{mykf}{Key Finding}%
{colback=green!5,colframe=green!35!black,fonttitle=\bfseries}{th}

\setcopyright{acmlicensed}
\copyrightyear{2025}
\acmYear{2025}
\acmDOI{XXXXXXX.XXXXXXX}

\acmJournal{CSUR}

\begin{document}
\newcommand{\ihsen}[1]{\marginpar{\color{blue}\tiny\ttfamily Ihsen: #1}}
\title{Emerging Threats and Countermeasures in Neuromorphic Systems: A Survey}


\author{Pablo Sorrentino}
\email{p.f.a.sorrentino@rug.nl}
\orcid{0009-0005-6009-2992}
\affiliation{%
  \institution{Faculty of Science and Engineering, University of Groningen}
  \city{Groningen}
  \country{Netherlands}
}

\author{Stjepan Picek}
\email{stjepan.picek@ru.nl}
\orcid{0000-0001-7509-4337}
\affiliation{%
  \institution{Faculty of Electrical Engineering and Computing, University of Zagreb}
  \city{Zagreb}
  \country{Croatia}
}
\affiliation{%
  \institution{Radboud University}
  \city{Nijmegen}
  \country{The Netherlands}
}

\author{Ihsen Alouani}
\email{i.alouani@qub.ac.uk}
\orcid{0000-0001-5102-8087} 
\affiliation{%
  \institution{Centre for Secure Information Technologies (CSIT), Queen's University Belfast}
  \city{Belfast}
  \country{United Kingdom}
}

\author{Nikolaos Athanasios Anagnostopoulos}
\email{Nikolaos.Anagnostopoulos@uni-passau.de}
\orcid{0000-0003-0243-8594} 
\affiliation{%
  \institution{Faculty of Computer Science and Mathematics, University of Passau}
  \city{Passau}
  \country{Germany}
}

\author{Francesco Regazzoni}
\email{f.regazzoni@uva.nl}
\orcid{0000-0001-6385-0780}
\affiliation{%
  \institution{University of Amsterdam}
  \city{Amsterdam}
  \country{The Netherlands}
}
\affiliation{%
  \institution{Università della Svizzera italiana}
  \city{Lugano}
  \country{Switzerland}
}

\author{Lejla Batina}
\email{lejla.batina@ru.nl}
\orcid{0000-0003-0727-3573}
\affiliation{%
  \institution{Digital Security (DiS) Group, Radboud University}
  \city{Nijmegen}
  \country{Netherlands}
}

\author{Tamalika Banerjee}
\email{t.banerjee@rug.nl}
\orcid{0000-0001-6848-0467}
\affiliation{%
  \institution{Faculty of Science and Engineering, University of Groningen}
  \city{Groningen}
  \country{Netherlands}
}

\author{Fatih Turkmen}
\email{f.turkmen@rug.nl}
\orcid{0000-0002-6262-4869}
\affiliation{%
  \institution{Faculty of Science and Engineering, University of Groningen}
  \city{Groningen}
  \country{Netherlands}
}

\renewcommand{\shortauthors}{Sorrentino et al.}

\begin{abstract}
Neuromorphic computing mimics brain-inspired mechanisms through spiking neurons and energy-efficient processing, offering a pathway to efficient in-memory computing (IMC). However, these advancements raise critical security and privacy concerns. As the adoption of bio-inspired architectures and memristive devices increases, so does the urgency to assess the vulnerability of these emerging technologies to hardware and software attacks.
Emerging neuromorphic computing architectures introduce new attack surfaces, particularly due to asynchronous, event-driven processing and stochastic device behavior. The integration of memristors into neuromorphic hardware and software implementations in spiking neural networks (SNNs) offers diverse possibilities for advanced computing architectures, including their role in security-aware applications. This survey systematically analyzes the security landscape of neuromorphic systems, covering side-channel and fault-injection attacks, hardware Trojans, adversarial and backdoor attacks, privacy attacks, and countermeasures. We also discuss security primitives enabled by neuromorphic and emerging-memory devices, including Physical Unclonable Functions (PUFs), True Random Number Generators (TRNGs), and secure in-memory or near-sensor computation. By mapping existing attacks, defenses, and security primitives in neuromorphic systems and examining their assumptions, validation methods, and implementation costs, the paper provides a basis for identifying practical limitations and developing secure and trustworthy neuromorphic architectures.
\end{abstract}

\ccsdesc[500]{General and reference~Surveys and overviews}
\ccsdesc[500]{Security and privacy~Hardware security implementation}
\ccsdesc[300]{Computer systems organization~Neural networks}
\ccsdesc[300]{Hardware~Emerging technologies}

\keywords{Neuromorphic Computing, Spiking Neural Networks, Side Channel Attacks, Physical Unclonable Functions, True Random Number Generators, Hardware Security, In-Memory Computing, Emerging Non-Volatile Memories}




\maketitle
\section{Introduction}




With the rise of data-intensive applications, traditional computing architectures are struggling to address their power consumption adequately. This issue is closely related to the \textit{von Neumann bottleneck} \cite{zou2021vonNeumann}, where the physical separation between memory and processing units results in significant energy and latency costs due to constant data movement. In response, neuromorphic computing has gained attention as a promising alternative, offering brain-inspired architectures that exploit parallelism and event-driven computation to provide significant reductions in energy consumption \cite{steven_2016_energy}. These advancements also introduce new and often underestimated security challenges. IMC represents a paradigm shift, enabling a wide range of implementations across various applications \cite{schuman_opportunities_2022}. In this context, neuromorphic and in-memory computing are closely related, both aiming to overcome the limitations of conventional architectures, as neuromorphic computing often utilizes in-memory computing as a key component of its architecture. 

However, as industry and academia advance with bio-inspired architectures and in-memory computing paradigms, ensuring the security and privacy of these emerging systems becomes increasingly important \cite{mehonic_brain-inspired_2022_} before widespread adoption. Some of the concrete security and reliability challenges include different exposure to side-channel attacks (SCAs), susceptibility to fault-injection attacks (FIAs), software- and hardware-level vulnerabilities, reliability issues in memory-based architectures \cite{al-meer_physical_2022}, and exploitability of memory footprint to extract sensitive information \cite{kocher_spectre_2019}. The intersection of these new paradigms with evolving attack methodologies underscores the need for a comprehensive understanding of their vulnerabilities and mitigation strategies.

In recent decades, research on software and hardware security has grown significantly, driven by concerns about data privacy and the resilience of traditional computing systems to sophisticated attacks \cite{gupta2025ai}. However, the security implications of neuromorphic systems remain less explored, particularly when software-level attacks interact with hardware substrates, device non-idealities, and in-memory computation. These challenges include hardware Trojans, watermarking, encryption, reverse engineering, and attacks that target diverse architectures or arise from cross-domain software interactions.

A flagship example of in-memory computing and neuromorphic computing is the integration of spiking neural networks (SNNs) with memristive devices. Memristor-based implementations of SNNs present unique security challenges that remain insufficiently addressed. In this survey paper, we present a thorough analysis of the vulnerabilities and countermeasures in neuromorphic computing, with particular attention to hardware and software approaches for integrating spiking neural networks with memristive devices and the security implications of Physical Unclonable Functions (PUFs) and True Random Number Generators (TRNGs). We examine attacks such as power and electromagnetic side-channels, and adversarial manipulations, while also discussing architectural and algorithmic defenses. By reviewing existing literature and reported experimental evidence, we provide a comprehensive perspective on the security landscape of neuromorphic computing. Additionally, we review the techniques that aim to strengthen the system's resistance to various attacks by implementing countermeasures like data obfuscation and adversarial training, while considering the energy and implementation costs that may affect their practicality.


To the best of our knowledge, no existing survey provides a unified perspective that links device-level behavior, neuromorphic architectures, and vulnerabilities within a common threat framework. This leaves critical gaps at the intersection of materials science, device physics, and security research. For instance, the security implications of material-specific phenomena—such as variability, noise, or retention—remain underexplored, and material scientists rarely analyze attack vectors. In contrast, software security research often ignores how emerging devices interact with system-level defenses or coexist with conventional computing platforms (e.g., hybrid memristor--CMOS architectures or mixed analog--digital systems). Furthermore, there is no integrated perspective that combines hardware, software, and architectural countermeasures into coherent co-design strategies. This study seeks to address these gaps by systematically examining vulnerabilities and defenses, highlighting both risks and opportunities, including the use of device variability for PUFs and TRNGs, and laying the groundwork for secure and resilient neuromorphic architectures. 


To support transparency and long-term usability, we maintain an accompanying open-source repository with curated references, structured comparison tables, and supplementary resources related to this survey.\footnote{The repository is available at: \href{https://github.com/SorrentinoPablo/Survey_Security_Privacy_Neuromorphic_Computing/}{Security and Privacy of Neuromorphic Computing}.}

The paper is organized as follows. Section II reviews prior work, identifying key contributions and research gaps in neuromorphic security. Section III outlines the research methodology and details the criteria used to select and analyze the relevant literature. Section IV explores neuromorphic hardware vulnerabilities and corresponding countermeasures, focusing on device-level threats and attack vectors. Section V addresses software-level vulnerabilities in neuromorphic systems, including adversarial attacks and mitigation techniques. Section VI examines both intrinsic and device-level security features of neuromorphic hardware. It includes traditional primitives such as Physical Unclonable Functions (PUFs) and True Random Number Generators (TRNGs), as well as broader mechanisms that leverage neuromorphic or in-memory architectures to combine processing, memory, and encryption within a single substrate. Section VII concludes the survey, highlighting open challenges, future research directions, and the integration of hardware-software perspectives toward secure neuromorphic architectures.

\section{Background}

This section first introduces the principles and motivations behind neuromorphic computing, highlighting its potential to overcome the limitations of traditional architectures. Spiking Neural Networks (SNNs) are then presented as a key neural-network model within these systems. Finally, the section outlines emerging security threats associated with neuromorphic systems, drawing on security lessons from traditional computing platforms, including vulnerabilities to side-channel attacks, fault-injection attacks, and hardware Trojans, as well as the role of hardware primitives. Together, these elements establish the conceptual and technological background necessary for the paper.

\subsection{Neuromorphic In-memory Computing}
\label{subsec:inmemory}

Neuromorphic computing represents a paradigm shift in computer architecture, inspired by the structure and operation of the human brain, and aims to address the limitations of traditional von Neumann designs \cite{indiveri_memory_2015}. In conventional systems, the physical separation between the memory and processing units presents a barrier regarding data processing speed and reduced energy consumption. Neuromorphic architectures mitigate this bottleneck by integrating computation into memory, enabling reduced data transfer with lower power and higher throughput. By mimicking the behavior of neurons and synapses through specialized hardware, these systems are well suited to workloads characterized by sparse, event-driven, and sensory data processing, including pattern recognition and machine learning tasks \cite{schuman_opportunities_2022}.

Spiking Neural Networks (SNNs) provide a computational model for neuromorphic systems by representing information through temporally sparse spike events. Their operation is based on neuron and synapse models that describe membrane-potential dynamics, spike generation, and activity-dependent changes in synaptic strength \cite{SNN_NeuroM_Rathi}. A key mechanism in this process is synaptic plasticity, where the strength of connections between neurons adapts based on prior activity. Mathematical models, including the Spike Response Model (SRM) and the Izhikevich model, use differential equations to describe membrane potential dynamics and spike generation. Neuromorphic computing relies on both continuous and discrete formulations. In discrete models, spike-timing-dependent plasticity (STDP) describes how synaptic weights evolve over time based on the spike timing between connected neurons \cite{jouni2025stdp}.

\begin{figure}[ht]
  \centering
  \includegraphics[width=0.60\linewidth]{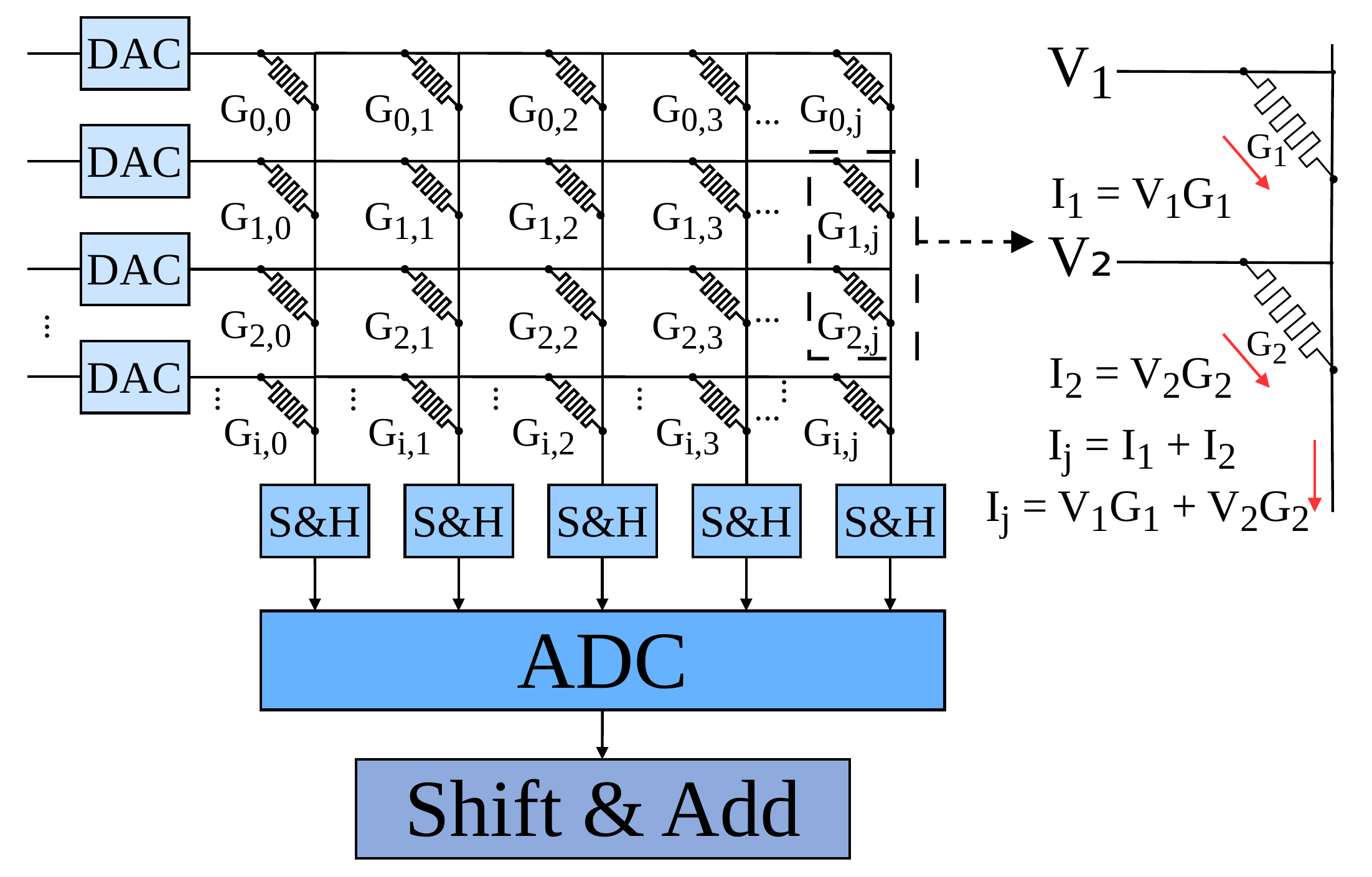}
  \caption{Memristor crossbar array performing matrix--vector multiplication, where input voltages and device conductances generate accumulated output currents. This illustrates the multiply-and-accumulate (MAC) operation used in many analog neuromorphic and in-memory computing implementations, where computation and storage are co-located within the same physical structure \cite{staudigl_survey_2022}.}
  \Description{Description}
  \label{fig:memristor_crossbar}
\end{figure}

In-Memory Computing (IMC) is a hardware approach that can support efficient implementations of Spiking Neural Networks in neuromorphic systems by reducing data movement between memory and processing units. This connection is particularly relevant in memristive architectures, where memory and computation can be co-located within the same physical substrate. Memristors, two-terminal devices whose resistance depends nonlinearly on the history of voltage and current, are frequently used to implement neuromorphic architectures \cite{SNN_NeuroM_Rathi}. Because of their ability to store multiple resistance states, memristors enable memory elements to emulate synaptic weights, allowing artificial neural networks to approximate the behavior of biological synapses. Figure~\ref{fig:memristor_crossbar} illustrates a memristor crossbar array performing matrix-vector multiplication, where input voltages interact with device conductance to produce output currents.

Memristors provide a hardware substrate for spike-based computation through their intrinsic non-volatility, nonlinear behavior, and ability to store multiple conductance levels \cite{peng_memristor-based_2024}. These characteristics enable them to emulate synaptic plasticity by adjusting connection strengths based on spike timing. As neural networks have evolved from perceptrons and sigmoid-based models to third-generation spiking neural networks, memristive devices have emerged as a core enabler of synapse-like behavior \cite{peng_memristor-based_2024}. However, training SNNs remains a challenge due to their non-differentiable, event-driven nature, which limits the applicability of standard backpropagation and often leads to higher training costs in terms of energy and time. To address these limitations, alternative strategies have been explored for real-time and energy-constrained applications, including local rule-based learning schemes and artificial-to-spiking neural network (ANN-to-SNN) conversion techniques. These training mechanisms play a key role in determining how effectively neuromorphic systems can support practical computation.

Memristors, both non-volatile and volatile, are central components in neuromorphic computing due to their ability to combine memory and processing within an energy-efficient framework. Their conductance modulation properties can support more linear and symmetric synaptic weight updates, which can improve the precision and stability of learning in SNNs. Beyond synaptic emulation, recent hardware demonstrations emphasize scalability, CMOS compatibility, and three-dimensional integration as requirements for practical memristive neuromorphic systems \cite{aguirre_hardware_2024}. These device-level advances also connect neuromorphic computing with probabilistic and brain-inspired models, where stochasticity and low-power operation become part of the computing substrate \cite{ham_neuromorphic_2021}. Also, detailed device–circuit simulation tools, such as Memristor-Spikelearn \cite{liu2023memristor}, demonstrate the importance of incorporating realistic memristor behavior when modeling synaptic plasticity and energy–accuracy trade-offs. From a security perspective, these same device properties are also relevant because variability, stochastic switching, retention effects, and conductance drift can shape attack surfaces, influence the effectiveness of defenses, and enable hardware-based security primitives \cite{oh2022memristor}. 

\subsection{Spiking Neural Networks}

\begin{figure*}[t]
  \centering
  \includegraphics[width=0.80\textwidth]{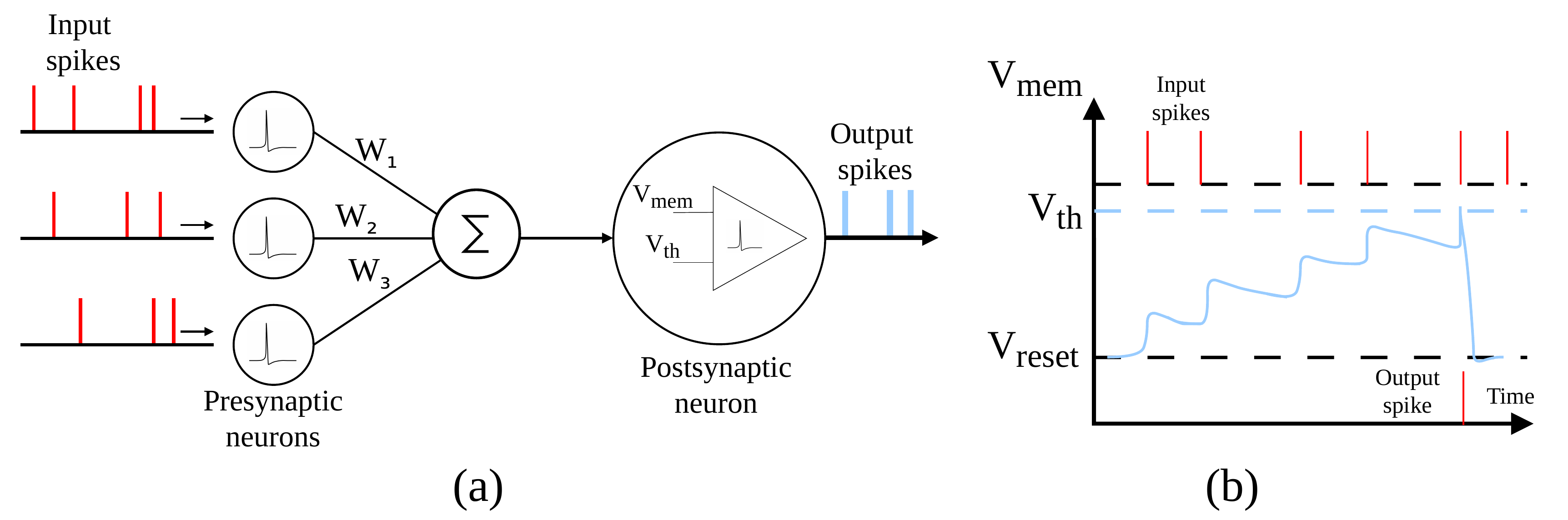}
  \caption{Overview of spike-based processing in SNNs. (a) Classical SNN connectivity, where presynaptic neurons transmit input spikes through weighted synapses to a postsynaptic neuron. (b) Leaky Integrate-and-Fire (LIF) neuron dynamics, where input spikes are integrated into the membrane potential until a threshold is reached, triggering an output spike and reset.}
  \Description{Two-panel schematic of spiking neural network computation. The first panel shows presynaptic neurons connected through weighted synapses to a postsynaptic neuron. The second panel shows the membrane potential of a leaky integrate-and-fire neuron increasing after input spikes, crossing a firing threshold, generating an output spike, and resetting.}
  \label{fig:SNN_overview}
\end{figure*}

Spiking Neural Networks (SNNs) process information through discrete spike events rather than the continuous activations used in traditional artificial neural networks (ANNs). This event-driven computation can reduce unnecessary activity and support temporal processing, making SNNs promising for low-power and time-dependent applications \cite{zendrikov2023brain}. Information in SNNs can be encoded through spike timing, spike rates, or population activity, depending on the coding scheme. Learning can be implemented through several mechanisms, including local rules such as Spike-Timing-Dependent Plasticity (STDP), surrogate-gradient methods, and artificial-to-spiking neural network (ANN-to-SNN) conversion. In STDP, synaptic weights are adjusted based on the relative timing of pre- and postsynaptic spikes: potentiation occurs when a presynaptic spike precedes a postsynaptic one, while depression occurs when the reverse happens. Figure~\ref{fig:SNN_overview} summarizes these concepts by showing both the connectivity of a classical SNN, where multiple presynaptic neurons influence a postsynaptic neuron, and the membrane-potential dynamics of a Leaky Integrate-and-Fire (LIF) neuron. Together, these figures illustrate how spike inputs are integrated over time and converted into output spikes when the membrane potential crosses a firing threshold.

The LIF model is a widely used spiking neuron model in deep SNN architectures, where neurons accumulate input spikes into a membrane potential that resets once a predefined threshold is exceeded \cite{hunsberger2015spiking}. This process is described using differential equations that capture the temporal evolution of the membrane potential and the conditions for spike generation. Compared to conventional neural models, this formulation introduces additional complexity due to its discrete-event nature and time dependency. The resulting threshold-based dynamics lie at the core of SNN behavior. Such mathematical modeling is essential for understanding the adaptability, stability, and learning capacity of spiking networks under various training rules \cite{ding_enhancing_2024}. By grounding spiking behavior in biologically inspired equations, these models enable the development of efficient neuromorphic architectures.

Implementing SNNs requires specialized simulation frameworks designed to accommodate their event-driven, time-dependent behavior. While conventional deep learning libraries such as TensorFlow and PyTorch are optimized for ANNs, SNN development is supported by simulators and training and conversion frameworks \cite{nunes2022spiking}. These tools provide the infrastructure to model spiking activity and manage discrete-event dynamics. SNN programming involves specifying the network topology (e.g., number of layers and neurons), the neuronal dynamics (such as membrane potential equations), and the synaptic learning rules (including Spike-Timing-Dependent Plasticity or other biologically inspired mechanisms). However, despite these advantages, SNN-based systems also introduce new challenges in terms of implementation complexity, model stability, and exposure to novel forms of attack. In the next section, we examine the specific vulnerabilities associated with SNNs and neuromorphic architectures from a hardware and software security perspective.

\subsection{Threats}
Neuromorphic computing, due to its analog and event-driven nature, is exposed to a diverse set of security threats that differ from conventional digital computing. These threats can originate from the physical properties of the hardware, the dynamic behavior of spiking neural networks, or even software vulnerabilities in learning models. This survey focuses mostly on attack vectors related to side-channel attacks (SCAs), fault injection attacks (FIAs), hardware Trojans (HTs), and adversarial, backdoor, or privacy-focused software attacks.

Side-channel attacks are security threats that exploit unintended information leakage often originating from the physical operation of hardware systems. Such leakage often occurs naturally, as systems emit observable signals, such as power usage, timing variations, or electromagnetic emissions, during computation \cite{goos_private_2003}. In some cases, certain low-level signals may be intentionally exposed to support performance monitoring or debugging. However, if these signals are exposed without adequate safeguards, they can pose additional security risks. All processing systems generate outputs that may correlate with internal operations \cite{brier_correlation_2004}, creating opportunities for attackers to infer sensitive data. Recognizing these forms of leakage is essential when evaluating the security of hardware platforms.


Hardware and software attacks can be further categorized based on the attacker’s interaction with the target. In passive attacks, adversaries observe leakage signals without interfering with the system. In contrast, active attacks introduce intentional faults through techniques like voltage manipulation or laser injection and analyze the system’s response to extract sensitive information. These attacks are also distinguished by the number of observations required: some techniques rely on a single power trace to retrieve information, while others employ multi-trace statistical analysis to uncover hidden patterns \cite{kocher1999differential}. These classifications shape the feasibility, cost, and stealthiness of attacks, especially in modern hardware and learning-based systems.


Designing effective defenses against hardware and software attacks requires a clear understanding of the adversary’s capabilities, knowledge, and access. This distinction is especially important in interdisciplinary neuromorphic research, where device-level studies often report variability, noise, or robustness without explicitly defining the attacker’s knowledge, access, or objective. The attack model is typically defined by the attacker’s level of system awareness and access. White-box adversaries possess full system knowledge, including source code and implementation details, allowing them to craft precise, targeted attacks. Gray-box adversaries have partial access—such as hardware specifications but not model parameters—while black-box attackers rely solely on external input-output behavior. These attacks may also be categorized based on physical access: remote attacks exploit emissions or timing from a distance, whereas local attacks involve direct hardware interaction. Advanced techniques often combine multiple data sources, including power traces, runtime patterns, and noise filtering.

Unlike broad taxonomies that merge heterogeneous labels into a single hierarchy, this framework provides a consistent basis for comparing the reviewed studies throughout the survey. In particular, the validation dimension explicitly distinguishes simulation-based studies from FPGA, ASIC, fabricated-chip, and physical-measurement evaluations.

\begin{figure}[ht]
  \centering
  \includegraphics[width=0.95\linewidth]{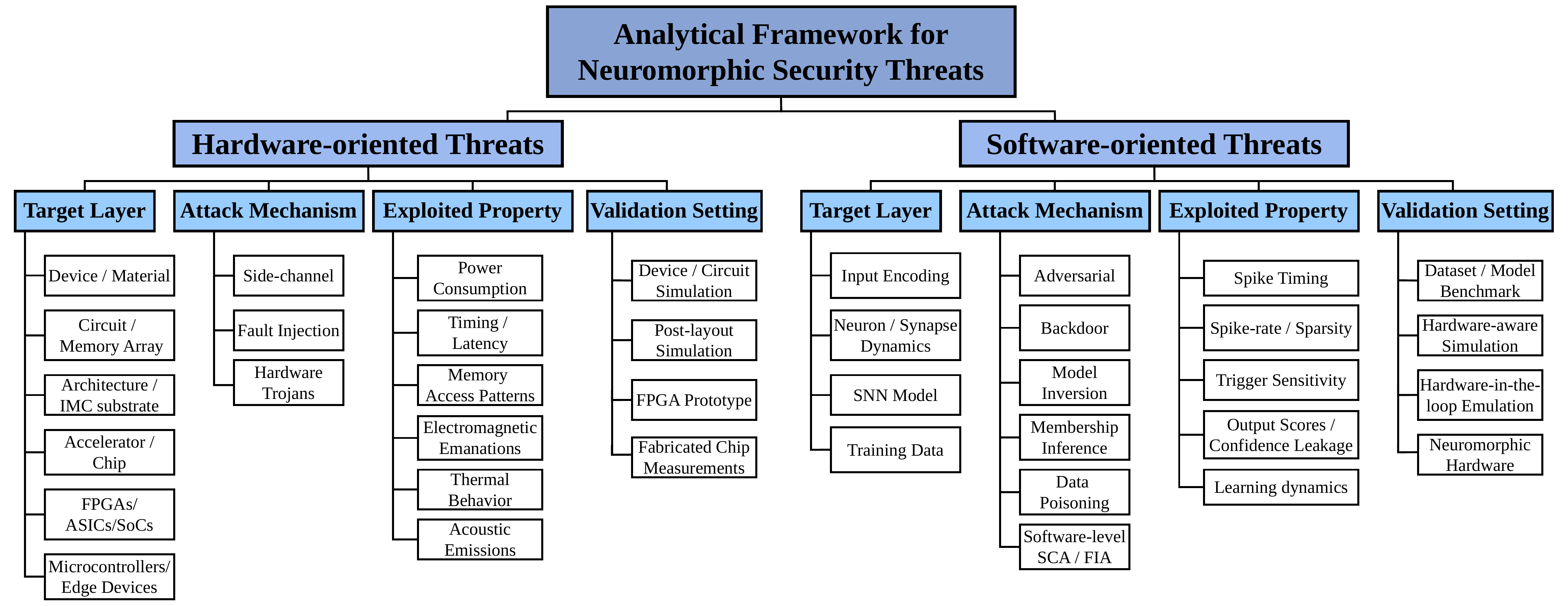}
  \caption{Cross-layer analysis framework used to classify neuromorphic security works in this paper. The framework separates the affected system layer, attack mechanism, exploited physical/computational property, and validation setting, including the distinction between simulation-based and physically validated studies.}
  \Description{Attacks}
  \label{fig:Attacks}
\end{figure}


\subsection{Review of Previous Survey Papers}

Previous reviews have focused on individual parts of the security landscape considered in this survey, including physical side-channel analysis, emerging-memory security, and adversarial robustness in neural networks. However, they do not jointly evaluate vulnerabilities and countermeasures across neuromorphic hardware, SNN models, and security enablers. Horvath et al.~\cite{horvath2024sok}, for example, systematically decompose physical side-channel attacks into phases, align them with threat models, and examine neural-network implementations. Their analysis provides a useful methodological basis for studying physical leakage, but it is not specific to event-driven neuromorphic systems or emerging-memory devices. Table~\ref{tab:survey-summary} presents a comparison between this paper and previous reviews with the closest conceptual or methodological overlap. Studies on conventional neural networks, general IMC, or peripheral memory systems are included only when they provide directly transferable evidence for a concrete neuromorphic attack surface.

The closest prior work is the survey by Staudigl et al.~\cite{staudigl_survey_2022}, which provides an early overview of neuromorphic computing-in-memory architectures, devices, simulators, and security considerations. Its security coverage is necessarily constrained by its broader architectural scope and publication date, with security discussed through a relatively small set of earlier studies. The analysis primarily addresses potential vulnerabilities in emerging memories, memristive crossbars, and neuromorphic Multi-Processor System-on-Chip (MPSoCs), while several attack classes examined in the present survey are absent or only briefly discussed. In particular, it does not provide a systematic treatment of physical side channels, fault injection, hardware Trojans, adversarial and backdoor attacks, privacy attacks, or security enablers such as PUFs and TRNGs. Its distinction between neuro attacks and neuro-CMOS attacks provides a useful initial separation, but it does not independently characterize the affected layer, attack mechanism, exploited property, attacker assumptions, and validation setting. Our survey extends this early perspective through an updated cross-layer analytical framework for neuromorphic and emerging-memory security.

A recent review focuses specifically on adversarial attacks and defense mechanisms in SNNs \cite{survey_AA}. It categorizes gradient-based and adjacent adversarial strategies, discusses structural and training factors affecting SNN robustness, and compares the susceptibility of SNNs and ANNs under these strategies. However, its scope is limited to adversarial robustness at the model level. It does not examine the attack types (e.g., backdoors), vulnerabilities, and security functionalities as we do in this paper. 
It also does not provide a cross-layer synthesis of validation maturity and implementation overhead across software and neuromorphic hardware.

Another recent survey examines the security landscape of embedded non-volatile memories, covering architectural vulnerabilities, side-channel and fault-injection attacks, information leakage, thermal threats, security primitives (e.g., PUF) and logic obfuscation \cite{Zakia_Survey} with a focus on emerging-memory technologies. 
However, its unit of analysis is the eNVM subsystem rather than the complete neuromorphic system. The paper does not examine spike-based computation, SNN-specific software attacks, temporal encoding, or the interaction between neuromorphic workloads and their physical implementation. Our survey complements this memory-centric perspective by connecting emerging-memory vulnerabilities and security primitives with SNN models, neuromorphic architectures, software-level attacks, and cross-layer validation settings. More generally, we provide a broader and more up-to-date coverage of scientific papers at the device, memory, and architectural levels. 


\begin{table*}[t]
  \caption{Comparison with the closest prior surveys.
  \(\checkmark\), \(\triangle\), and -- denote systematic, partial,
  and no coverage, respectively.}
  \label{tab:survey-summary}
  \footnotesize
  \centering
  \setlength{\tabcolsep}{6pt}
  \renewcommand{\arraystretch}{1.12}
  \begin{tabular}{@{}p{0.5cm} p{5.5cm} c c c@{}}
    \toprule
    \textbf{Work} &
    \textbf{Primary scope} &
    \textbf{HW security} &
    \textbf{SW security} &
    \textbf{Security enablers} \\
    \midrule

    \cite{staudigl_survey_2022}
    & Neuromorphic architectures and simulation frameworks, with limited security coverage
    & \(\triangle\)
    & \(\triangle\)
    & -- \\

    \cite{survey_AA}
    & Adversarial attacks and defense mechanisms in SNNs
    & --
    & \(\triangle\)
    & -- \\

    \cite{Zakia_Survey}
    & Security threats and security primitives in embedded non-volatile memories
    & \(\checkmark\)
    & --
    & \(\checkmark\) \\

    \textbf{Ours}
    & \textbf{Cross-layer security and privacy in neuromorphic systems}
    & \(\checkmark\)
    & \(\checkmark\)
    & \(\checkmark\) \\

    \bottomrule
  \end{tabular}
\end{table*}


\section{Research Methodology}

This survey adopts a structured and replicable methodology to explore the evolving landscape of security threats and countermeasures in neuromorphic computing, with a specific focus on side-channel attacks, fault-injection attacks, hardware Trojans, and software-level attacks on spiking neural networks. The methodology covers the planning, identification, selection, and analysis of relevant scientific literature published between January 2014 and June 2026. The review incorporates keyword refinement, boolean search logic, and inclusion/exclusion filtering.

The classification used in this survey emerged from an iterative literature-mapping process. During the initial step, papers were clustered according to their technical scope and type of contribution: hardware-oriented attacks, software-oriented attacks, and countermeasures. A third category covered hardware-based security primitives and security-enabling mechanisms, including PUFs, TRNGs, and related device- or circuit-level security features. These works were grouped together because they exploit physical variability, stochastic behavior, or hardware-level co-design to support entropy generation, uniqueness, protection, or secure computation. This grouping enables a consistent analysis of these mechanisms as \textit{security enablers} within neuromorphic and emerging-memory architectures, rather than as isolated techniques.

\subsection{Research Questions}
To guide this systematic review, we formulate the following five research questions (RQs), each targeting a specific dimension of the security landscape in neuromorphic computing systems:
\begin{itemize}
    \item \textbf{RQ1}: \textit{What are the emerging trends, publication patterns, and research gaps in security for neuromorphic computing?}
    
    \item \textbf{RQ2}: \textit{What are the main hardware-level attacks against neuromorphic computing systems such as side-channel attacks and fault-injection attacks, and what methodologies are used to evaluate them?}
    
    \item \textbf{RQ3}: \textit{How do software-level attacks, including adversarial, backdoor, and privacy attacks, target SNN-based systems, and which vulnerabilities are  exploited to perform these attacks?}
    
    \item \textbf{RQ4}: \textit{What countermeasures have been proposed against hardware- and software-level threats in neuromorphic systems, and how do their energy, area, latency, or accuracy costs affect their practicality?}
    
    \item \textbf{RQ5}: \textit{How are hardware-based security primitives and security-enabling mechanisms, including PUFs, TRNGs, and reliability-oriented techniques, integrated into neuromorphic and emerging-memory architectures?}
    
\end{itemize}

\subsection{Search Strategy}
The search process was conducted using a structured search string divided into three keyword groups (see Table~\ref{tab:taxonomy-scope}): \textit{Fundamental}, \textit{Scope}, and \textit{Context}. Boolean operators and truncation symbols were employed to improve precision and coverage. The queries were applied to the title, abstract, and keyword fields of the retrieved records, followed by a review of methodology and conclusions to verify relevance. 

To cover relevant papers, each search string included one or more keywords from the Fundamental group, one from the Scope group, and at least one from the Context group. The first layer defined the overall domain of security and computing paradigms, the second identified the specific neuromorphic or in-memory hardware contexts, and the third captured concrete threat types, defense mechanisms, and security applications. This approach allowed consistent filtering of studies addressing the intersection between hardware and software security in neuromorphic systems.

\begin{table*}
  \caption{Overview of the keyword groups considered during the data collection phase, categorized into fundamental terms, scope-related terms, and contextual terms.}
  \label{tab:taxonomy-scope}
  \small
  \begin{tabular}{p{2cm} p{10.5cm}}
    \toprule
    Group & Keywords \\
    \midrule
    Fundamental & Hardware Attacks, Software Attacks, Security Modules, Applications, In-Memory Computing Platforms \\[6pt]
    
    Scope & Neuromorphic Chip, In-Memory Computing, Spiking Neural Network, AI Accelerator, Memristor/ReRAM/NVM/RRAM/FeFET/Spintronic Devices, Crossbar Array, Neuromorphic Sensor, Physical Unclonable Function, True Random Number Generator, Hardware Security Module \\[6pt]
    
    Context & Side-channel Attack/Analysis, Power SCA, Electromagnetic SCA, Timing Attack, Thermal Attack, Fault Injection Attack, Trojan Attack, Row Hammer Attack, Threat, Logic Locking, Data Poisoning, Watermark, Backdoor Attack, Membership Inference Attack, Machine Learning Guided Attack, Software Attack, Adversarial Attack, Evasion Attack, Attacks in Deep Neural Networks, Reverse Engineering, Model Inversion Attack, Robustness, Countermeasure/Defense/Mitigation \\
    \bottomrule
  \end{tabular}
\end{table*}
The search targeted high-impact digital libraries relevant to neuromorphic computing, hardware and software security, and machine learning, reflecting the growing interest in neuromorphic security. Studies were selected based on their type of contribution and relevance to the topic, while duplicate entries and non-accessible papers were excluded.

\begin{figure}[ht]
  \centering
\includegraphics[width=0.75\linewidth]{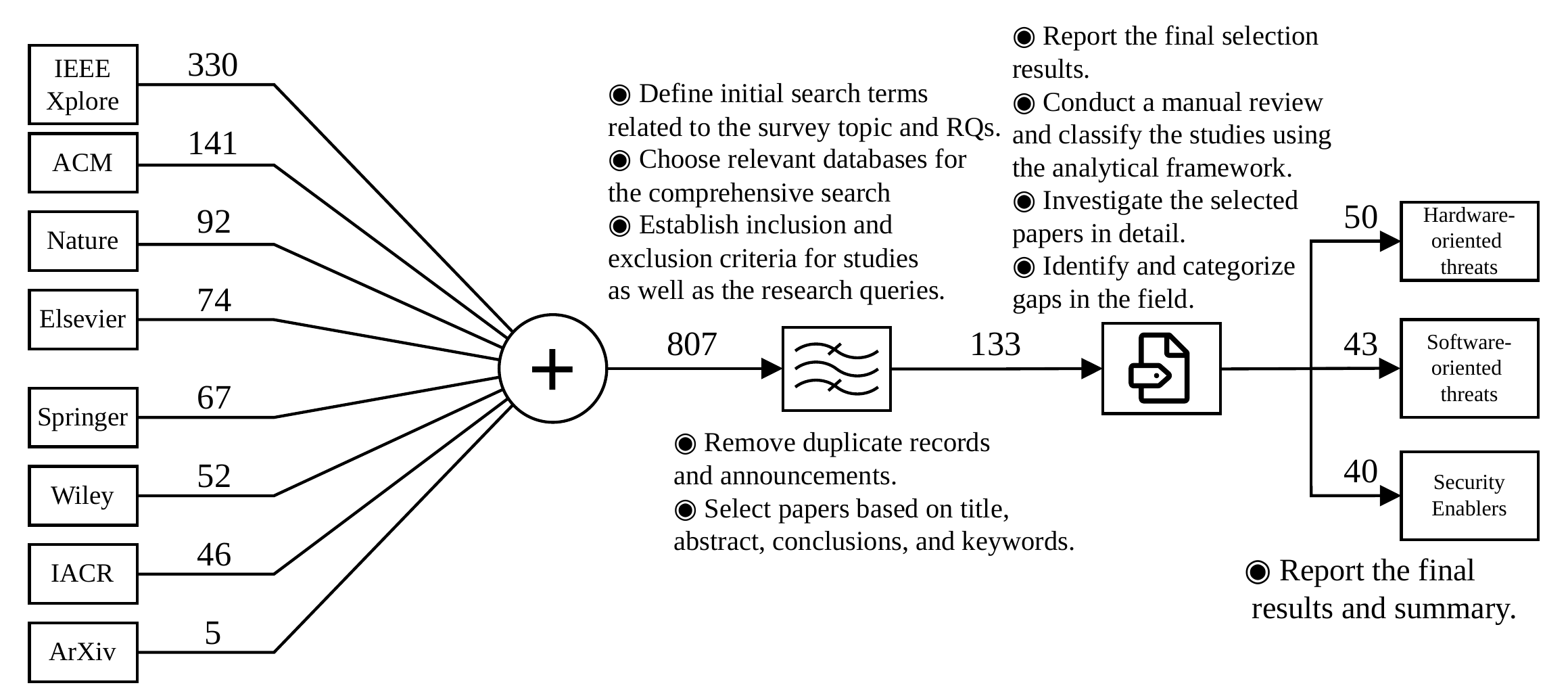}
  \caption{Illustration of the selection methodology of this survey.}
  \label{fig:study-design1}
  \Description{Design process.}
\end{figure}

To ensure consistency in the analysis, the selected articles were organized according to the analytical framework proposed, which was designed to align with the previously defined research questions. Each article was manually mapped to one or more relevant sections and subsections, based on its focus on attack mechanisms, defensive strategies, validation settings, or security-enabling architectural features. This categorization enables a systematic and focused examination of the literature, facilitating clearer comparisons and thematic insights.

The hierarchical structure of the literature classification is illustrated in Figure~\ref{fig:neuromorphic_security_map}, showing its division into topic areas and the number of articles assigned to each category. To avoid conflating different dimensions, the classification distinguishes between the attack mechanism, the target layer, the exploited property, and the validation setting. For example, side-channel and fault-injection attacks are treated as attack mechanisms, while their impact may be analyzed at the device, circuit, architecture, or model level depending on the study. While studies categorized in the hardware domain primarily focus on physical mechanisms of information leakage or fault injection (e.g., power traces, electromagnetic emissions, or laser-induced faults), those assigned to the software domain analyze how these attacks affect neural models or learning behavior (e.g., spiking activity distortion or weight manipulation). This distinction was intentionally adopted to encompass both perspectives.
\begin{figure*}[t]
    \centering
    \tikzset{
      main/.style    = {rectangle, draw, rounded corners, ultra thick,
                        align=center, minimum height=1.2em, text width=6em, fill=gray!5, scale=0.7},
      section/.style = {rectangle, draw, rounded corners, very thick,
                        align=center, minimum height=1.0em, text width=5.0em, fill=gray!10, scale=0.7},
      sub/.style     = {rectangle, draw, rounded corners,
                        align=center, minimum height=1.0em, text width=4.2em, fill=gray!5, scale=0.7},
      refs/.style    = {rectangle, draw, rounded corners, dashed,
                        align=center, minimum height=1.0em, text width=4.2em, fill=white, inner sep=1.5pt, scale=0.7},
      arrow/.style   = {very thin,->}
    }

    \begin{tikzpicture}[node distance=0.35cm and 2.6cm, every node/.style={font=\scriptsize}]

      \node[main] (main) at (0,0) {\textbf{Security\\Privacy in\\Neuromorphic Devices}};

      \node[section, below=0.35cm of main, xshift=-6.8cm] (secH) {\textbf{Hardware\\Security}};
      \node[section, below=0.35cm of main]                (secS) {\textbf{Software\\Security}};
      \node[section, below=0.35cm of main, xshift= 6.8cm] (secA) {\textbf{Security\\Enablers}};

      \draw[arrow] (main) -- (secH);
      \draw[arrow] (main) -- (secS);
      \draw[arrow] (main) -- (secA);

      \node[sub, below=0.30cm of secH, xshift=-2.0cm] (h1) {SCA};
      \node[sub, below=0.30cm of secH]                (h2) {FIA};
      \node[sub, below=0.30cm of secH, xshift= 2.0cm] (h3) {HT};

      \node[refs, below=0.18cm of h1] (h1r) {\cite{iranfar2025power} \cite{jiang2020mempoline} \cite{bommana2026phantom} \cite{rakin_deepsteal_2021} \cite{zou_enhancing_2022} \cite{kubota_deep_2021} \cite{wang_side-channel_2024} \cite{shao_imce_2024} \cite{mir2024extracting} \cite{adel_robust_2024} \cite{oh2022memristor} \cite{gupta2025ai} \cite{aji2026rethinking} \cite{chakraborty_correlation_2017} \cite{roy2024security} \cite{inglese2023side} \cite{iranfar_layout-oriented_2024} \cite{li_protfe_2024} \cite{masoumi_novel_2020} \cite{merkel2025power} \cite{ensan_scare_2020} \cite{wang2023powergan} \cite{yan_mercury_2023} \cite{sapui2024side} \cite{yuan_hypertheft_2024} };
      \node[refs, below=0.18cm of h2] (h2r) {\cite{wang2024cimsat} \cite{divyanshu_logic_2022} \cite{du2021low} \cite{parrini_error_2024} \cite{querlioz_simulation_2011} \cite{rezayati2025new} \cite{wang2023nvleak} \cite{chien_attack_2023} \cite{jouni2025stdp} \cite{huber2025write} \cite{xu2021situ} \cite{staudigl2023fault}  \cite{ashok2025digital} \cite{venceslai_neuroattack_2020} \cite{lv2022variation} \cite{yang2019thwarting} \cite{mahmod_untrustzone_2024} \cite{spyrou2021neuron} \cite{staudigl_neurohammer_2022} \cite{staudigl_nvm-flip_2024} \cite{inglese2023side} \cite{heidary2024hardware}
      };
      \node[refs, below=0.18cm of h3] (h3r) {\cite{srinivas_era_2024} \cite{rakin_tbt_2020} \cite{rajamanikkam_understanding_2021} \cite{wu_hardware_2023} \cite{raptis2025input}};

      \draw[arrow] (secH) -- (h1); \draw[arrow] (h1) -- (h1r);
      \draw[arrow] (secH) -- (h2); \draw[arrow] (h2) -- (h2r);
      \draw[arrow] (secH) -- (h3); \draw[arrow] (h3) -- (h3r);

      \node[sub, below=0.30cm of secS, xshift=-2.8cm] (s1) {Adversarial};
      \node[sub, below=0.30cm of secS, xshift=-1.0cm] (s2) {SCA/FIA};
      \node[sub, below=0.30cm of secS, xshift= 1.0cm] (s3) {Backdoor};
      \node[sub, below=0.30cm of secS, xshift= 2.8cm] (s4) {Inference};

      \node[refs, below=0.18cm of s1] (s1r) {\cite{lammie2025inherent} \cite{el2021securing} \cite{ding_snnrat_2022} \cite{chen2025robust} \cite{du2025raw} \cite{he_parametric_2019} \cite{kundu_hire-snn_2021} \cite{nomura2022robustness} \cite{siddique_moving_2024} \cite{yu2026time} \cite{liang2021exploring} \cite{liang_toward_2022} \cite{rosenberg_2022} \cite{wu2023threshold} \cite{xu_attacking_2025} \cite{yadav2024unveiling} \cite{ahmad2023security} \cite{yao2024exploring} \cite{chen2025power} \cite{lun2025towards}};
      \node[refs, below=0.18cm of s2] (s2r) {\cite{venceslai_neuroattack_2020} \cite{zou_enhancing_2022} \cite{batina_csi_nodate} \cite{siddique_improving_2023} \cite{yan2025adversarial} \cite{spyrou2021neuron} \cite{jouni2025stdp} \cite{wong_snngx_2024} \cite{bu2023rate} \cite{krithivasan_efficiency_2022} \cite{nazari_securing_2024}};
      \node[refs, below=0.18cm of s3] (s3r) {\cite{abad_sneaky_2024} \cite{fu_spikewhisper_2024} \cite{rakin_tbt_2020} \cite{zhang2024event} \cite{poursiami_watermarking_2024} \cite{adi2018turning} \cite{marchisio_is_2020} \cite{poursiami2025spikes}};
      \node[refs, below=0.18cm of s4] (s4r) {\cite{li2024membership} \cite{poursiami_watermarking_2024} \cite{moshruba2024neuromorphic}};

      \draw[arrow] (secS) -- (s1); \draw[arrow] (s1) -- (s1r);
      \draw[arrow] (secS) -- (s2); \draw[arrow] (s2) -- (s2r);
      \draw[arrow] (secS) -- (s3); \draw[arrow] (s3) -- (s3r);
      \draw[arrow] (secS) -- (s4); \draw[arrow] (s4) -- (s4r);

      \node[sub, below=0.30cm of secA, xshift=-2.0cm, yshift=-0.04cm] (a1) {PUFs};
      \node[sub, below=0.30cm of secA,               yshift=-0.04cm] (a2) {TRNGs};
      \node[sub, below=0.30cm of secA, xshift= 2.0cm, yshift=-0.04cm] (a3) {Features};

      \node[refs, below=0.18cm of a1] (a1r) {\cite{wali_hardware_2023} \cite{yue2025physical} \cite{adel_robust_2024} \cite{oh2022memristor} \cite{florian_using_2022} \cite{james2025processing} \cite{zhao2024understanding} \cite{ibrahim_resilience_2024} \cite{jeon2019physical} \cite{john_halide_2021} \cite{al-meer_physical_2022} \cite{zhang2025enhancing} \cite{zhang2025enhancing} \cite{Zhang2025PUF}};
      \node[refs, below=0.18cm of a2] (a2r) {\cite{chien_attack_2023} \cite{dodda_all--one_2022} \cite{hu_situ_2024} \cite{taneja_-memory_2022} \cite{abunahla_memristor_2018} \cite{wali_hardware_2023} \cite{bahador_algorithmically-enhanced_2024} \cite{bende2026rram} \cite{shao_imce_2024} \cite{zhu2024neuromorphic} \cite{wang2025monolithic} };
      \node[refs, below=0.18cm of a3] (a3r) {\cite{bahador2026novel} \cite{dodda_all--one_2022} \cite{hu_situ_2024}  \cite{lammie2025inherent} \cite{divyanshu_logic_2022} \cite{du2021low} \cite{ashok2025digital} \cite{shao_imce_2024} \cite{mir2024extracting} \cite{chen2025protected} \cite{zhu2024neuromorphic} \cite{wang2026high} \cite{xu2023embedding} \cite{park2026model} \cite{sun2025cmos} \cite{ding2025transforming}};

      \draw[arrow] (secA) -- (a1); \draw[arrow] (a1) -- (a1r);
      \draw[arrow] (secA) -- (a2); \draw[arrow] (a2) -- (a2r);
      \draw[arrow] (secA) -- (a3); \draw[arrow] (a3) -- (a3r);

    \end{tikzpicture}

    \caption{Structure of the survey organizing selected papers into main sections and subsections based on attack types, countermeasures, and security applications.}
    \label{fig:neuromorphic_security_map}
    \Description{Papers per topic.}
\end{figure*}

Figure~\ref{fig:neuromorphic_security_map} complements the last representation by detailing the number of studies assigned to each section and subsection. These results show that neuromorphic security research is diversifying across hardware attacks, software attacks, and security-enabling mechanisms, forming a coherent but still maturing research landscape. The following sections build on this distribution to analyze each dimension in depth.

\section{Neuromorphic Hardware: Vulnerabilities and Countermeasures}

\begin{figure}[ht]
  \centering
  \includegraphics[width=0.8\linewidth]{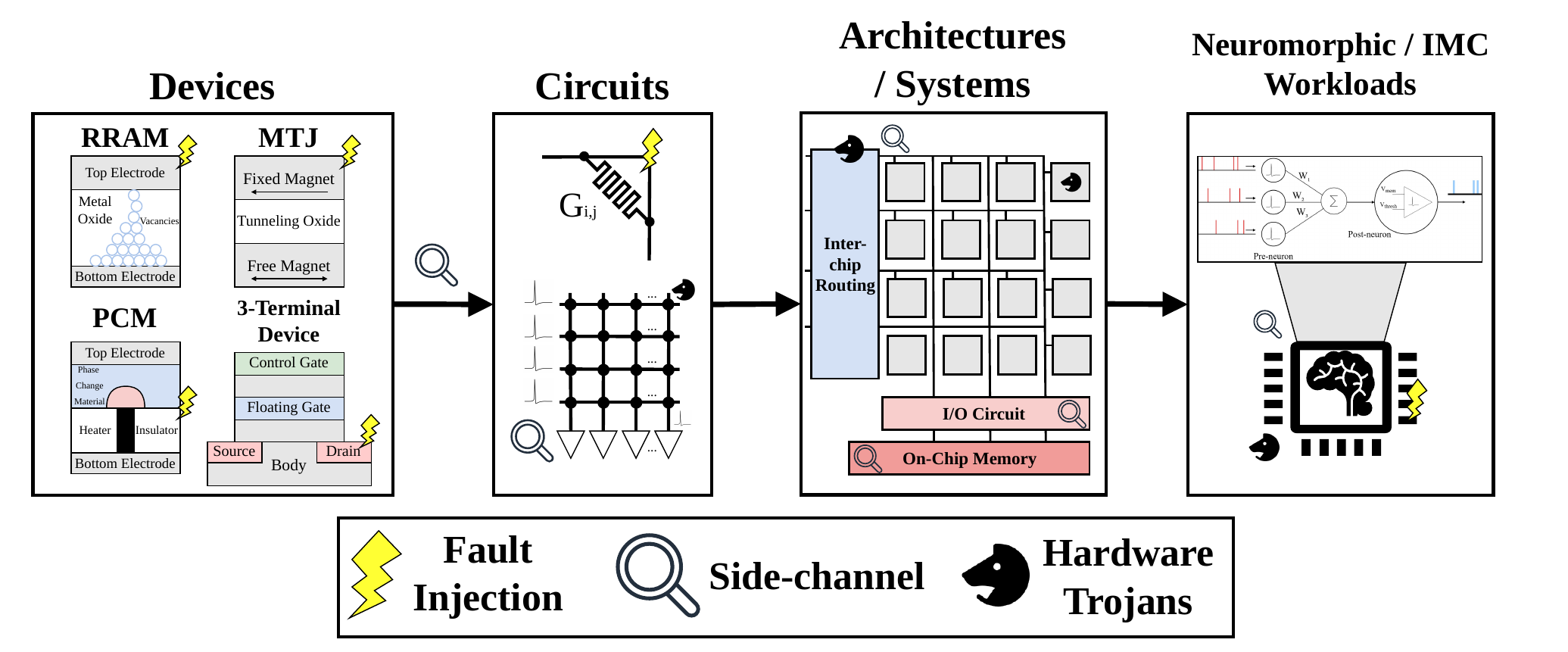}
  \caption{Architectural overview of neuromorphic computing hardware, from individual devices to in-memory computing architectures. Attack vectors such as side-channel attacks, fault injection, and hardware Trojans are mapped onto the architectural layers \cite{nazari_securing_2024}.}
  \label{fig:study-HWA}
  \Description{Layered schematic of neuromorphic hardware showing devices, memory arrays, peripheral circuits, and in-memory computing architectures, with side-channel attacks, fault injection, and hardware Trojans mapped to the corresponding hardware layers.}
\end{figure}

This section examines the main categories of hardware-level attacks targeting neuromorphic computing platforms, focusing on side-channel leakage, fault-injection techniques, and malicious hardware modifications. As neuromorphic architectures move from simulations toward actual device-, circuit-, and system-level implementations, understanding their exposure to physical-layer threats  is crucial. Figure~\ref{fig:study-HWA} provides an architectural overview that maps attack vectors to different hardware layers, from individual devices and memory arrays to full in-memory computing systems. The analysis that follows is structured into four parts: (i) side-channel attacks, including power, electromagnetic, timing, and thermal leakage; (ii) fault-injection attacks, including perturbations that affect device states, memory behavior, or neural computation; (iii) hardware Trojans and malicious circuit-level modifications; and (iv) countermeasures aimed at improving robustness through detection, error correction, redundancy, and fault tolerance. Each category is examined in terms of attack mechanisms, validation setting, potential impact, and possible countermeasures.

\subsection{Side-channel Attacks}

Side-channel attacks exploit information unintentionally leaked through the physical execution of a system, including power consumption, electromagnetic emissions, timing behavior, and thermal variations. In conventional platforms, these attacks are often distinguished by the adversary’s access to profiling data, ranging from profiled attacks that use a reference behaviour to non-profiled attacks that rely only on observations from the target system \cite{picek_sok_2023}. In neuromorphic systems, however, side-channel leakage is shaped by event-driven computation, sparse spike activity, asynchronous communication, and memory-centric processing. Although these properties are usually introduced to reduce average power consumption and improve energy efficiency, they can also produce input-dependent physical signatures. In this setting, spike timing, firing rates, memory access patterns, and crossbar current variations may correlate with the processed data, model parameters, or neural activity. As a result, adversaries may exploit power, electromagnetic, timing, or thermal measurements to infer sensitive information about neuromorphic workloads, deployed models, or hardware states.

\subsubsection{Power Side-channel Attacks}

Power side-channel analysis has emerged as a significant threat to neuromorphic accelerators and in-memory computing architectures. A first study \cite{wang_side-channel_2024} showed that, under specific conditions, power traces collected from resistive RAM-based computing platforms can be used to reverse-engineer neural network models without accessing stored weights. By analyzing bit-serial input operations, attackers were able to infer key structural elements, such as layer ordering, output channel widths, and convolutional kernel sizes, using only dynamic power profiles. The analysis relied on a hardware-aware mixed-signal power simulator, rather than physical IMC measurements, and was validated by reconstructing a LeNet model. The simulator incorporates component-level power models and evaluates sampling-rate and noise effects, but the attack remains unvalidated on fabricated RRAM-based IMC hardware. These vulnerabilities may extend to event-driven architectures like IBM’s TrueNorth \cite{akopyan2015truenorth} and Intel’s Loihi \cite{lin2018programming}, where spike-based activity can produce traceable power signatures. Although not designed for neuromorphic hardware, the MERCURY framework \cite{yan_mercury_2023} is relevant here because it demonstrates an automated remote power side-channel attack against an off-the-shelf NVIDIA DNN accelerator. MERCURY collects voltage-fluctuation traces during model inference and formulates architecture recovery as a sequence-to-sequence learning problem, recovering the victim model topology with an error rate below 1\%. Unlike simulation-only IMC studies, MERCURY uses measured time traces from an FPGA-based NVIDIA prototype, providing stronger empirical evidence for remote power leakage. These results expose privacy and intellectual-property concerns for accelerator-based neural inference and motivate similar evaluations in neuromorphic hardware.

Several studies have highlighted the vulnerability of neuromorphic and memory-based systems to physical side-channel leakage. Prior work on correlation power analysis has shown that data-dependent switching activity in specialized low-power hardware can produce exploitable power variations, suggesting that similar leakage mechanisms may arise in neuromorphic processors where computation is distributed across devices, memory arrays, and communication fabrics \cite{chakraborty_correlation_2017}. More directly, Merkel and Su \cite{merkel2025power} show that power profiles in NVM crossbar-based neuromorphic systems can leak information related to loss sensitivity, enabling adversaries to craft evasion attacks even without any knowledge of the training dataset. Their study demonstrates attacks on both single-layer and multi-layer neural implementations and shows that Bayesian optimization can reduce the query cost, although the results remain simulation-based and require validation under real crossbar non-idealities. 

Complementing this attack perspective, Iranfar et al. \cite{iranfar2025power} propose a hybrid MTJ/CMOS AES accelerator for Processing-in-Memory (PiM) architectures with symmetric power behavior designed to resist Differential Power Analysis (DPA) or Correlation Power Analysis (CPA). Post-layout and system-level simulations show constant power-consumption patterns under process variations, reduced correlation between secret-dependent operations and observable power traces, and improved static power and operating frequency compared with conventional ASIC and FPGA baselines. These works show that power traces are both a leakage source and a design constraint in emerging memory-based computing, motivating hardware validation of side-channel risks and countermeasures under realistic device conditions. Beyond leakage-based attacks, recent work shows that CiM security defenses can also be vulnerable to formal de-obfuscation and reverse-engineering attacks. CiMSAT \cite{wang2024cimsat} targets hardware obfuscation schemes in CiM architectures by combining functional approximation of storage behavior with a bias-tolerant SAT attack for mixed-signal computation. The study evaluates 176 defense configurations derived from 14 CiM obfuscation techniques and reports that 90\% can be de-obfuscated within 1,000 seconds, while 98\% are recovered within approximately one day. After recovering the obfuscation keys, the reconstructed CiM models recover high classification accuracy on MNIST and CIFAR-10, showing that static-key obfuscation alone is insufficient for protecting stored weights and inference behavior. This result raises the bar for CiM and neuromorphic security: defenses must be evaluated not only against physical leakage, but also against algorithmic de-obfuscation attacks that exploit the mathematical structure of in-memory computation.

Neuromorphic and memory-based systems are also vulnerable to power-based side-channel attacks, where variations in power consumption may reveal internal states, data-dependent operations, or architectural features. Chen et al. \cite{chen2025power} address this problem from a memristive cryptographic-circuit perspective by proposing a power-balanced hiding strategy based on memristor groups. Their approach aims to equalize the power cost of logic operations during writing and reading cycles, reducing the dependence between logic inputs and observable power traces. 

Architecture-level countermeasures have also been proposed to reduce side-channel leakage in neuromorphic and logic-in-memory systems. In \cite{masoumi_novel_2020}, a hybrid CMOS/memristor implementation of the AES algorithm was shown to resist DPA by leveraging the intrinsic non-linear and multilevel behavior of memristors. These characteristics introduced less predictable power consumption profiles, reducing the correlation between processed data and leakage patterns. Using circuit simulations and DPA models in MATLAB, Synopsys, HSPICE, and Cadence Spectre, the study reported improved DPA resistance and lower simulated energy consumption for the hybrid CMOS/memristor design. However, physical measurements were performed only for the CMOS baseline, while the memristive implementation remained simulation-based. 

Similarly, the vulnerability of circuits to power side-channel leakage has been examined in \cite{iranfar_layout-oriented_2024}. These architectures are often proposed for neuromorphic systems due to their non-volatility and energy efficiency, yet their resistance to DPA and CPA remains uncertain. Using Monte Carlo simulations and process-variation analysis, the study evaluated logic gates and memory blocks under varying conditions. The results showed that symmetric layouts, particularly those using Pre-Charge Sense Amplifiers (PCSAs), achieved the lowest leakage. In contrast, asymmetric designs exhibited high data-dependent variation. These findings underscore the importance of layout-aware countermeasures, such as power balancing and symmetry enforcement, to secure architectures deployed in neuromorphic and cryptographic applications.

Overall, existing work shows that power side-channel leakage is a credible concern for neuromorphic, in-memory, and emerging-memory architectures, but the strength of the evidence remains uncertain. FPGA-based studies provide actual measured traces, while many IMC, memristive, and NVM-crossbar analyses remain simulation studies. Post-layout and variation-aware studies offer stronger circuit-level evidence, particularly when they report area, delay, and power. However, fabricated-chip measurements under realistic device variability, noise, temperature, and workload conditions remain limited. This gap is important because several proposed defenses reduce leakage only by introducing trade-offs in power, latency, mapping complexity, or hardware overhead.

\subsection{Fault Injection Attacks}

While side-channel attacks extract information from unintentional leakage, fault injection attacks deliberately induce errors in hardware behavior. In neuromorphic computing, these faults often exploit the sensitivity of memristive elements and storage states to electrical stress. NeuroHammer \cite{staudigl_neurohammer_2022} is a notable example that adapts the RowHammer principle to memristor-based memory by exploiting thermal crosstalk between adjacent memory cells. The attack is evaluated through a device- and circuit-level simulation flow combining COMSOL Multiphysics and Cadence Virtuoso to model thermal crosstalk in dense ReRAM crossbars. Although this provides hardware-aware evidence, the evaluation remains simulation-based and uses a deterministic memristive model, without physical crossbar measurements. In the modeled setting, such bit flips could corrupt stored weights, memory states, or inference-related data by targeting specific cell positions. The study demonstrates that a 70\% bit flip rate can be achieved in targeted cells with fewer than 10,000 access cycles, suggesting the feasibility of access-driven fault injection in non-volatile memory. While these simulations demonstrate feasibility, their physical accuracy depends strongly on device geometry, material properties, and thermal coupling, and thus the results should be interpreted as indicative rather than conclusive evidence.

Beyond direct thermal-crosstalk-based fault injection, related emerging-memory threats can also arise from access-pattern effects, read-time asymmetries, aging, and retention behavior. Although these mechanisms are not always fault-injection attacks in the strict sense, they are relevant to neuromorphic hardware because they exploit device- and memory-level physical behavior that can corrupt, reveal, or bias computation. Staudigl et al. \cite{staudigl_nvm-flip_2024} extend the implications of NeuroHammer from isolated ReRAM crossbar cells to the system level by introducing NVgem5, a gem5-based simulator for injecting NeuroHammer-induced bit flips into eNVM cache and main-memory models. Their results show that a single cache-level bit flip can compromise RSA signature generation and leak secret-key information, highlighting how device-level faults can propagate into system-level security failures. However, the evaluation remains architectural and simulation-based, with abstracted bit-flip thresholds and patterns rather than physical eNVM hardware validation.

A related, but distinct, class of memory-level attacks exploits data-dependent access latency rather than directly inducing faults. Chowdhury et al. \cite{chowdhuryy_understanding_2023} introduce R-SAW, an attack framework that exploits read-time asymmetries in multi-level phase-change memory (PCM). Through memory access patterns and execution-cycle analysis, the attack extracts sensitive data from the distribution of information across storage lines. Because this is primarily a timing side-channel rather than a fault-injection attack, its relevance to this section lies in showing that emerging-memory devices can expose security risks through physical read behavior even without explicit bit-flip induction. These findings indicate that neuromorphic systems relying on multi-level memories require memory designs and access protocols that account for both disturbance-induced faults and data-dependent read behavior.

Fault-injection attacks can disrupt neuron behavior, spike propagation, or classification accuracy in neuromorphic systems. Although direct attack demonstrations on fabricated neuromorphic platforms remain limited, injected faults could plausibly alter neuron activation, corrupt intermediate spike representations, or affect synchronization-dependent mechanisms. As a hardware-oriented countermeasure to fault-induced errors, Spyrou et al. \cite{spyrou2021neuron} propose a neuron fault-tolerance strategy for SNNs based on large-scale fault-injection experiments. Their approach first uses dropout during training to make some neuron fault types passive, and then applies active fault detection and recovery for the remaining faults. For hidden layers, the authors introduce offline and online detection schemes, together with a fault-hopping mechanism that simplifies recovery; and for the output layer, they use triple modular redundancy. The strategy is evaluated on convolutional SNNs using N-MNIST and DVS128 Gesture, showing that neuron-level detection and recovery can improve resilience in benchmark SNNs. However, the evaluation is behavioral and software-based, and the exact area, power, and latency costs of a complete hardware implementation require further validation.

A related class of threats exploits long-term data remanence and aging effects in on-chip SRAM. Mahmod et al. \cite{mahmod_untrustzone_2024} show that data-dependent transistor aging can imprint secrets into SRAM cells and later reveal them through power-on state measurements across multiple commodity devices. Although this attack does not target neuromorphic hardware directly, it is relevant to neuromorphic processors that rely on SRAM caches or peripheral memories, highlighting aging and retention effects as possible sources of long-term information leakage.

Emerging non-volatile memories, including magnetic tunnel junction (MTJ)-based designs, can also introduce fabrication-time security risks in neuromorphic and in-memory computing systems. Chowdhury et al. \cite{chowdhury_s-tune_2024} show that small manufacturing-level modifications, such as variations in insulating-layer thickness, can induce selective bit-flip behavior and increase classification errors for specific MNIST digits while leaving others mostly unaffected. As mitigation, the authors propose stochastic tuning and process-variation modeling to detect or reduce the impact of such manipulations. These results show that fabrication-time faults can create selective inference failures, making process-aware validation and material-level security analysis important for emerging neuromorphic hardware.

Overall, the fault-injection literature suggests that neuromorphic and emerging-memory systems are vulnerable to disturbance, aging, retention, and fabrication-induced faults, but the level of evidence remains uneven. NeuroHammer provides device-aware simulation of thermal crosstalk, while NVM-Flip evaluates the system-level consequences of NeuroHammer-induced bit flips through architectural simulation. Access-driven PCM attacks are mainly architectural side-channel studies, and neuron-level fault-tolerance studies are largely benchmark-based. Fabricated neuromorphic demonstrations remain limited, and the cost of defenses such as redundancy, monitoring, stochastic tuning, or error correction is not always quantified. Future work should therefore validate these attacks and countermeasures under realistic device variability, endurance, temperature, and workload conditions.

\subsection{Hardware Trojans}

The threat of Hardware Trojans (HTs) in neuromorphic and in-memory computing remains less explored than side-channel and fault-injection attacks. HTs can be inserted during design, layout, or fabrication, and triggered by rare input patterns, spike sequences, voltage conditions, or internal control states. Once activated, they may leak model parameters, suppress neurons, alter thresholds, or corrupt spiking activity while remaining hidden under normal stochastic behavior. This risk is amplified in analog and emerging-memory systems, where device variability, drift, and noise can mask malicious effects. Wu et al.~\cite{wu_hardware_2023} demonstrate this threat through a supply-chain Trojan against analog eNVM neural accelerators, where a Trojan payload suppresses selected neuron ADCs and uses transient power signatures to recover more than 90\% of synaptic weights. However, the evaluation remains simulation-based, relying on NeuroSim and transistor-level circuit analysis rather than fabricated eNVM hardware.

Related studies show that Trojan behavior in neural systems can also emerge beyond conventional malicious circuitry. Meyers et al.~\cite{meyers2024trained} show that maliciously trained weights can increase the correlation between FPGA power traces and neural-network outputs, enabling remote power side-channel recovery of classification results without adding hardware resources. Although this is not a traditional HT and does not target neuromorphic or emerging-memory hardware directly, it highlights a cross-layer risk in which model parameters themselves can amplify hardware leakage. Raptis et al.~\cite{raptis2025input} further explore Trojan behavior in SNNs by proposing an input-gated Trojan neuron activated by engineered spiking patterns with minimal hardware overhead. Overall, these works suggest that HTs in neuromorphic systems may exploit not only hidden circuits, but also peripheral activity, model parameters, and spike-triggered dynamics. Systematic evaluations, detection benchmarks, and fabricated-platform demonstrations remain limited.

\subsection{Fortifying the Hardware}

As side-channel and fault injection attacks continue to expose vulnerabilities in neuromorphic hardware, several countermeasures have been proposed at the circuit and architecture levels. Li et al. \cite{li_protfe_2024} introduced ProtFe, a power side-channel protection scheme for FeFET-based memories that conceals write traces by distributing current and introducing intermediate states. Importantly, this work explicitly quantifies the security--efficiency trade-off, reporting nearly zero area and latency overhead with 0.2\% energy overhead for PiMWrite, and 0.6\% area overhead with 7.1\% energy overhead. Defenses, from memory obfuscation to logic-level partitioning, demonstrate the importance of integrating hardware security throughout neuromorphic system design while accounting for energy consumption, area overhead, and implementation complexity.

Error resilience and fault tolerance help neuromorphic systems maintain correct behavior by detecting and correcting faults caused by noise, manufacturing variations, or attacks that could otherwise disrupt learning or be exploited by adversaries. To ensure computational reliability, \cite{parrini_error_2024} proposes a fault-tolerant architecture using synchronized checksum-based error detection and correction. Targeting FeFET and RRAM-based systems, the method integrates two redundant checksum blocks into a memory crossbar array. These units enable low-overhead, real-time arithmetic error detection and correction, achieving over 90\% recovery of model accuracy with moderate latency overhead. Dynamic correction embedded in neuromorphic cores can improve fault resilience while limiting, but not eliminating, efficiency costs.

Overall, the hardware literature shows that neuromorphic security risks are increasingly understood across devices, memory arrays, and accelerator architectures, but the maturity of the evidence remains insufficient. Based on the available evidence, power side-channel leakage currently represents the most empirically supported physical threat, particularly in FPGA-based SNN implementations and conventional accelerator platforms. In contrast, fault-injection and hardware-Trojan threats on fabricated neuromorphic hardware remain less mature and are supported mainly by device-aware simulation, behavioral evaluation, or isolated prototypes. Among countermeasures, balanced write and layout strategies with explicitly quantified overheads, together with targeted neuron-level fault-tolerance mechanisms, appear closest to practical deployment. Broader obfuscation, redundancy, and model-level defenses remain promising, but their scalability across device technologies and their complete hardware costs require further validation.

Finally, future efforts should prioritize hardware evaluation of side-channel and fault injection effects, and adopt layout-aware co-design frameworks that balance security, energy consumption, area overhead, and functionality. Table~\ref{tab:attack-summary} summarizes representative studies, including their attack or defense mechanisms, validation evidence, reported overhead, and main limitations.

\begin{table*}[t]
  \caption{Representative hardware attacks and defenses in
  neuromorphic and emerging-memory systems.
  N/R denotes not reported; N/Q denotes discussed but not quantified.}
  \label{tab:attack-summary}
  \centering
  \scriptsize
  \setlength{\tabcolsep}{2.5pt}
  \renewcommand{\arraystretch}{1.03}

  \begin{tabular}{@{}
    >{\raggedright\arraybackslash}p{0.95cm}
    >{\raggedright\arraybackslash}p{3.35cm}
    >{\raggedright\arraybackslash}p{2.30cm}
    >{\raggedright\arraybackslash}p{3.65cm}
    >{\raggedright\arraybackslash}p{3.20cm}
  @{}}
    \toprule
    \textbf{Paper} &
    \textbf{Role and mechanism} &
    \textbf{Target} &
    \textbf{Evidence} &
    \textbf{Cost / limitation} \\
    \midrule

    \multicolumn{5}{@{}l}{\textit{\underline{Side-channel attacks and defenses}}} \\
    \addlinespace[1pt]

    \cite{wang_side-channel_2024}
    & Power/timing architecture extraction
    & Mixed-signal RRAM IMC
    & 28-nm circuit-characterized simulation; full LeNet recovery under noise/sampling sweeps
    & Defense power/latency cost N/Q \\

    \cite{roy2024security}
    & ADC-output and activation leakage
    & MRAM IMC
    & Prototype and simulation
    & Partially characterized \\

    \cite{wang2023powergan}
    & GAN reconstruction from power traces
    & Analog CIM accelerator
    & Hardware-aware simulation; MRI recovery at 20\% power noise
    & Countermeasures tested; cost N/Q \\

    \cite{li_protfe_2024}
    & \textit{Defense:} multi-step writes and split-array balancing
    & FeFET memory arrays
    & Circuit simulation; 21$\times$/33$\times$ search-space expansion
    & PiMWrite: 0.2\% energy, near-zero area/latency; SpA: 0.6\% area, 7.1\% energy \\

    \cite{mir2026securing}
    & \textit{Defense:} balanced obfuscation and glitch-aware blinding
    & Digital SRAM-CIM accelerator
    & 40-nm gate-level simulation; CPA, TVLA, and clustering
    & CM-I: 44\% area/89.8\% energy; CM-II: 214\% area/356\% energy \\

    \addlinespace[1pt]
    \multicolumn{5}{@{}l}{\textit{\underline{Fault injection and fault tolerance}}} \\
    \addlinespace[1pt]

    \cite{staudigl_neurohammer_2022}
    & Repeated-write thermal crosstalk
    & ReRAM crossbar
    & COMSOL/Cadence VCM simulation; pulse, spacing, and temperature sweeps
    & Repeated writes required; no physical validation \\

    \cite{staudigl_nvm-flip_2024}
    & System-level hammer bit flips
    & ReRAM crossbar
    & NVgem5; $>40$k writes induce a memory fault and one cache fault recovers an RSA key
    & Requires adjacency/co-location; defense cost N/Q \\

    \cite{spyrou2021neuron}
    & \textit{Defense:} dropout, detection, fault hopping, and output TMR
    & Convolutional SNN neurons
    & SLAYER/PyTorch fault injection with transistor-derived models and 0.35-$\mu$m circuit simulation
    & Dropout: no HW cost; TMR covers only 0.57\%/0.04\% of neurons \\

    \addlinespace[1pt]
    \multicolumn{5}{@{}l}{\textit{\underline{Hardware Trojan attacks}}} \\
    \addlinespace[1pt]

    \cite{wu_hardware_2023}
    & ADC suppression and FFT-based power analysis
    & Analog eNVM neuromorphic accelerator
    & NeuroSim and 65-nm transistor simulation with PVT/noise; $>90\%$ weight recovery
    & Trojan area $<0.5\%$; layout masking about 10\% power/area \\

    \cite{raptis2025input}
    & Spike-pattern trigger and neuron saturation
    & Analog/digital SNN accelerators
    & Transistor analysis, and ZCU104 FPGA validation
    & 0.138\% LUT and 0.028\% power \\

    \addlinespace[1pt]
    \multicolumn{5}{@{}l}{\textit{\underline{Reverse-engineering attacks}}} \\
    \addlinespace[1pt]

    \cite{wang2024cimsat}
    & Bias-tolerant SAT de-obfuscation
    & Mixed-signal CiM
    & Algorithmic evaluation
    & Attack runtime reported \\

    \bottomrule
  \end{tabular}
\end{table*}

\section{Neuromorphic Software: Vulnerabilities and Countermeasures}

Software-level attacks against neuromorphic systems exploit the way spiking models encode, process, and expose information through spike timing, firing rates, and event-driven responses. While previous sections focused on physical vulnerabilities in memory and circuit components, equally important threats emerge from how software controls and interacts with these architectures. Spiking neural networks are a prime example of neuromorphic data processing. Because they process information through discrete electrical events, they introduce new behavior that can be exploited through crafted inputs or runtime manipulation. Their timing, energy use, and response patterns depend on spike generation, which adversaries can influence to degrade performance or extract sensitive information. These risks vary across implementation platforms, but all expose software-level vulnerabilities shaped by spike-based computation. Understanding how system behavior changes under adversarial control is essential for anticipating broader software threats.

The software layer, including the learning rules, model representations, and input encodings, often lacks standardized defensive strategies against manipulation. Attacks may exploit spiking frequency, memory state retention, or signal timing to compromise integrity or confidentiality. In this section, we examine how neuromorphic models are vulnerable to a wide range of software-level threats, including backdoor attacks, adversarial perturbations, membership inference, model inversion, and timing-aware manipulations. We also discuss countermeasures such as adversarial training, monitoring strategies, and robust design principles tailored to the neuromorphic computing domain.

\subsection{Threats in Software}

\subsubsection{Backdoor Attacks}
A major threat to neuromorphic computing arises from backdoor attacks, particularly in spiking neural networks, where spatio-temporal patterns can be embedded into training data to serve as covert triggers \cite{abad_sneaky_2024, li2025unsupervised}. Unlike traditional neural networks that process static inputs, SNNs respond to the timing and frequency of spikes, making them vulnerable to subtle manipulations that alter neuron activation without degrading general accuracy. When a specific spike sequence is used as a trigger, the model may misclassify the input only when that trigger is present while behaving normally on clean samples. These backdoors bypass several classic defenses based on static features, thus requiring temporal detection methods. SpikeWhisper \cite{fu_spikewhisper_2024}, for instance, spreads multiple triggers over time slots to evade detection in federated neuromorphic learning. Tests on standard neuromorphic datasets show high attack success rates with minimal impact on task performance. This work demonstrates that backdoors in SNNs pose a growing challenge that requires temporally aware defense mechanisms beyond those used in conventional deep learning. Complementary research by Sun et al. \cite{sun2024neural} introduces SODA, a causality-based framework that detects and mitigates semantic backdoors by tracing neuron-level cause–effect relationships in model decisions. While demonstrated on deep neural networks, this approach suggests that similar causal tracing could help identify covert temporal triggers in spiking systems.

\subsubsection{Adversarial Attacks}
\begin{figure}[ht]
  \centering
  \includegraphics[width=0.65\linewidth]{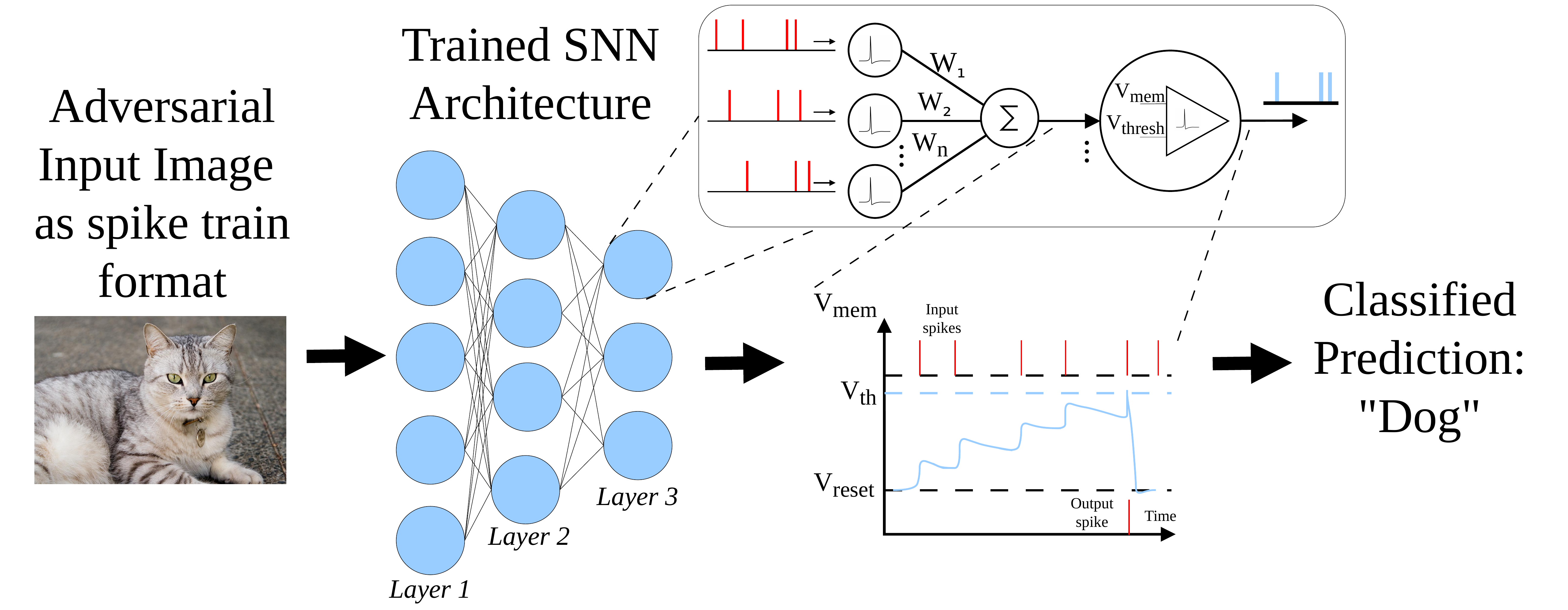}
  \caption{Schematic example of an adversarial attack against an SNN. A spatio-temporal perturbation modifies the spike-encoded input and causes the trained network to produce an incorrect classification.}
  \label{fig:study-SW1}
  \Description{Diagram showing an input image converted into a spike train, modified by an adversarial perturbation, processed by a multilayer spiking neural network, and assigned an incorrect output class.}
\end{figure}

By subtly altering the timing of spikes, attackers can mislead spiking neural networks, exploiting vulnerabilities similar to those seen in traditional adversarial attacks. In visual tasks, adversarial examples are often generated by gradient-based methods that modify pixel values; in SNNs, attackers can achieve a comparable effect by manipulating the temporal structure of spike trains. For example, shifting the spike time within a time-encoded input can alter the network’s output without changing the overall input semantics. Figure~\ref{fig:study-SW1} illustrates this mechanism by showing how small input perturbations can alter the predicted class in an SNN. Because SNNs process data dynamically, traditional techniques such as the Fast Gradient Sign Method (FGSM) \cite{xu_attacking_2025} must be adapted to account for temporal encoding rather than static feature space. This temporal sensitivity introduces unique vulnerabilities not present in conventional architectures and requires tailored methods to both generate and defend against adversarial examples in neuromorphic systems \cite{yao2024exploring}.

Beyond inducing misclassification, adversaries can also compromise the computational efficiency of spiking neural networks by manipulating their internal spiking behavior. Krithivasan et al. \cite{krithivasan_efficiency_2022} analyze how small input perturbations can degrade SNN performance by increasing spike activity, which results in higher energy consumption and longer execution times. Although SNNs are designed for energy-efficient and event-driven processing, the study introduces SpikeAttack, a method that takes advantage of this characteristic to disrupt performance with minimal changes to the input. Higher spike rates can translate into increased energy consumption and latency, depending on the execution platform and event-processing architecture. The authors show that such attacks are difficult to detect, making them especially concerning when SNNs are used under strict energy or timing constraints. However, the magnitude of this effect depends on the encoding scheme, firing threshold, and inference window; configurations that enforce sparse activity may reduce the attack impact.

\subsubsection{Software targets: Side-channel Attacks and Fault Injections}
\begin{figure}[ht]
  \centering
  \includegraphics[width=0.75\linewidth]{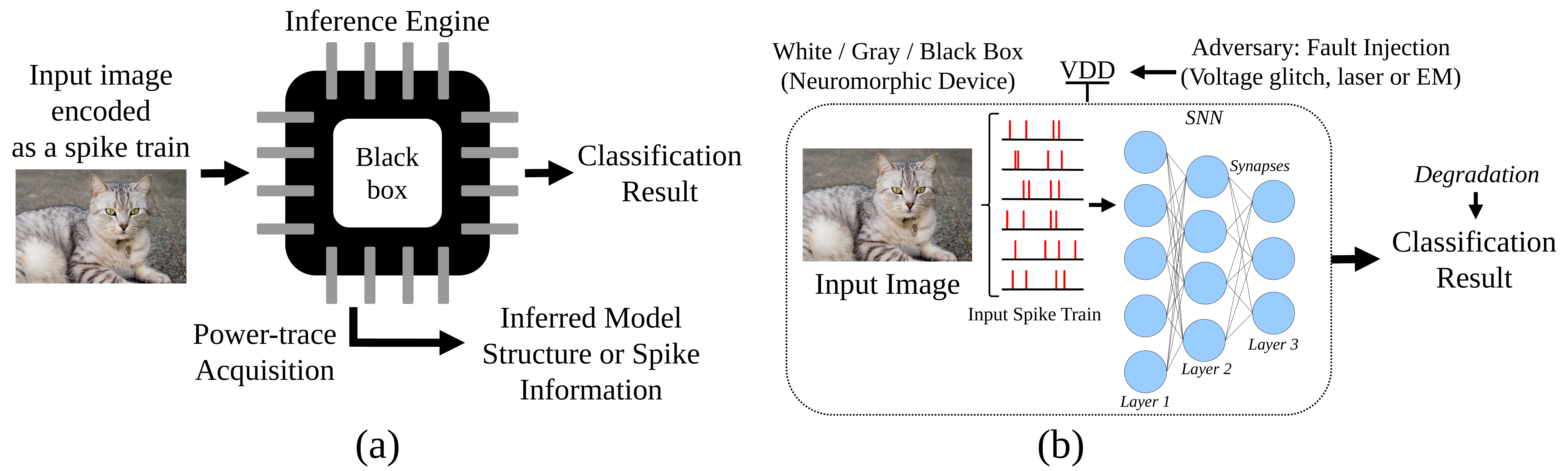}
  \caption{Cross-layer attacks affecting SNN execution. (a) A side-channel observer collects power traces to infer model structure or spike-related activity. (b) A fault-injection adversary perturbs the neuromorphic device, altering spike processing and degrading the classification result.}
  \label{fig:study-SW2}
  \Description{Two-panel diagram of cross-layer attacks on an SNN. The first panel shows power traces collected from an inference engine to recover neural-network structure. The second shows an adversary injecting a fault into a neuromorphic device and causing degraded classification.}
\end{figure}

Another critical vulnerability in neuromorphic systems involves side-channel and fault-injection attacks that affect software-level behavior. Rather than modifying the input or model directly, these attacks infer internal activity or perturb neural execution by observing external physical signals, such as fluctuations in power consumption. In SNNs, where computations are triggered by spikes and vary over time, attackers can correlate power traces with specific neural events to reconstruct sensitive information. Figure~\ref{fig:study-SW2} illustrates a representative example of such an attack, showing how a side-channel observer can extract timing or spike-related features from a neuromorphic chip executing an SNN model.

Recent work by Goswami et al. \cite{goswami2024investigation} demonstrates that power-based side-channel analysis can expose sensitive parameters in SNNs implemented on FPGAs, including neuron types like Izhikevich and Leaky Integrate-and-Fire. By applying differential power analysis on a two-layer network and later extending it to large-scale convolutional SNNs trained on FashionMNIST, the study shows that spiking activity correlates directly with measurable power variations. These variations reveal not only the number of firing neurons, but also characteristics linked to synaptic weight polarity and excitation patterns. Crucially, attackers can exploit this information to infer internal model behavior by deliberately selecting inputs that maximize energy consumption. These findings highlight the need for dedicated side-channel countermeasures adapted to the unique timing and sparsity properties of SNNs.

Nazari et al. \cite{nazari_securing_2024} analyze the vulnerabilities of on-chip learning in spiking neural networks, with a focus on security threats arising during training and inference on neuromorphic hardware. In scenarios where weight updates occur locally on the chip, adversaries may exploit access to data paths, power profiles, or neuron activation patterns. The study identifies specific attacks such as data poisoning, where malicious samples distort synaptic updates; evasion attacks, through crafted temporal inputs that bypass detection; and fault injection, which modifies membrane thresholds or connection strengths to influence learning outcomes. Furthermore, the paper highlights the feasibility of side-channel attacks that extract spike activity or synaptic behavior by monitoring dynamic power consumption during training phases. To counter these risks, the authors propose fault-aware mapping, noise-injected encoding to obfuscate power signatures, and redundant driver logic to limit side-channel leakage. These techniques, combined with adversarial training and input validation, form a tailored defense strategy for protecting neuromorphic systems engaged in real-time, on-chip learning.

CSI \cite{batina_csi_nodate} uses electromagnetic emissions to recover structural information from neural networks running on low-power ARM and AVR devices. Although evaluated on conventional DNNs rather than SNNs, the study provides adjacent evidence that event-driven neural accelerators may also expose model-dependent EM signatures. Also, the Targeted Bit Trojan attack \cite{rakin_tbt_2020} modifies a small number of model-weight bits to embed a hidden trigger that causes targeted misclassification while preserving normal behavior on benign inputs. Although demonstrated on conventional neural networks, it illustrates how post-deployment memory faults could introduce persistent and stealthy behavioral changes in SNN models.

Recent works by Nagarajan et al. \cite{nagarajan_analysis_2022, nagarajan_fault_2022}  reveal that spiking neural networks are vulnerable to fault injection attacks that exploit the system’s sensitivity to power fluctuations. By causing brief drops in voltage or using laser attacks, attackers can modify neuron parameters such as spike amplitude and firing threshold, leading to incorrect predictions during inference. These attacks are especially effective in low-power SNN hardware, where energy constraints limit the effectiveness of built-in protection. Experiments on digit recognition tasks \cite{nagarajan_analysis_2022} show that classification accuracy can be sharply reduced through controlled power anomalies, even without access to training data or internal architecture. The studies highlight how adversaries can exploit physical-level weaknesses to influence spike behavior and force predictable system errors. This line of work underscores the growing relevance of energy-based fault attacks in neuromorphic computing and the need for more robust, context-aware defenses that address the unique temporal and event-driven dynamics of SNN implementations \cite{nagarajan_fault_2022}.

\subsubsection{Privacy Attacks: Membership Inference and Model Inversion}

Membership inference attacks (MIAs), which aim to determine whether a specific data point was part of a model’s training set, have been extensively studied in artificial neural networks, but less so in spiking neural networks \cite{li2024membership}. Moshruba et al. \cite{moshruba2024neuromorphic} examine the privacy-preserving properties of spiking neural networks compared to conventional artificial neural networks, focusing on their resilience to these attacks. Through experiments on datasets such as MNIST, CIFAR-10, CIFAR-100, and ImageNet, they show that SNNs consistently achieve lower AUC ROC scores than ANNs, indicating improved privacy. The authors attribute the lower leakage observed in their experiments to spike-based activations and event-driven communication, which can make the input--output relationship less directly exploitable by the considered membership-inference attacks. Furthermore, SNNs trained with evolutionary algorithms achieve strong privacy resistance, reaching an AUC of 0.50 across all tests. The study also explores the trade-off between privacy and accuracy, showing that SNNs experience significantly less accuracy degradation than ANNs under differentially private stochastic gradient descent. 

Moreover, BrainLeaks' work \cite{poursiami2024brainleaks} presents a systematic study of model inversion attacks on spiking neural networks (SNNs), investigating how much private information can be reconstructed from trained models through their outputs. Unlike previous inversion techniques tailored for artificial neural networks (ANNs), BrainLeaks adapts the attack to the spiking domain by combining surrogate-gradient-based reconstruction with probabilistic modeling of spike trains to estimate class-distinctive features. Evaluations on both static (MNIST, AT\&T Face) and neuromorphic datasets (N-MNIST, IBM DVS128 Gesture) show that SNNs, while slightly more resilient than ANNs, still leak recognizable input patterns, disproving the assumption that non-differentiability inherently ensures privacy. These results highlight that the temporal and event-driven nature of SNNs does not fully protect against inversion and that the design of privacy-preserving mechanisms must extend beyond architectural differences. To mitigate this leakage, follow-up works suggest incorporating differentially private training frameworks. These findings position SNNs as a relevant platform for privacy-focused neuromorphic research, but not as inherently privacy-preserving models.

Taken together, these studies do not support treating SNNs as inherently privacy-preserving. Membership-inference results suggest reduced leakage relative to ANNs under some training and attack settings, whereas BrainLeaks demonstrates that recognizable information can still be reconstructed through SNN-specific inversion techniques. Current evidence remains predominantly benchmark-based, and privacy leakage has not yet been systematically evaluated across deployed neuromorphic hardware, different spike encodings, and realistic query constraints.



\subsection{Defense Strategies}


\subsubsection{Robust Training}
Training-time defenses mitigate adversarial manipulation in SNNs. One widely used technique involves injecting corrupted spike sequences—designed to mimic adversarial perturbations—into the training data \cite{el2021securing}. Additionally, real-time monitoring can help identify attack-induced changes during operation. By introducing redundant neuron channels and comparing outputs across them, SNN-based systems can detect inconsistencies that signal tampering or faults.

Beyond robustness during execution, protecting the intellectual property (IP) of SNN models is increasingly important. As examined in \cite{poursiami_watermarking_2024}, watermarking strategies originally designed for artificial neural networks—such as fingerprint-based and backdoor-based methods—have been adapted to the temporal characteristics of SNNs. While fingerprinting encodes identifiable patterns into parameters, backdoor-based watermarking embeds hidden triggers that can later validate ownership. The lower overparameterization in SNNs changes the balance, as they may be less affected by structural pruning but more sensitive to fine-tuning, which can disable weak labels. This trade-off underscores the need for watermarking methods tailored to the sparse and event-driven nature of SNNs. Future work must develop neuromorphic-aware IP protection that ensures ownership verification without compromising accuracy or computational efficiency.

\subsubsection{Regularization}
To enhance the security of neuromorphic systems without modifying hardware, MemPoline \cite{jiang2020mempoline} introduces a software-based defense that dynamically shuffles sensitive data in memory using parameter-guided permutations. This reordering limits an attacker's ability to infer secret values from memory access patterns, improving resistance against side-channel attacks while maintaining system performance. Originally designed for cryptographic algorithms like AES and RSA, this technique shows potential for protecting machine learning workloads, where memory-access regularity can expose model or data-dependent behavior. In parallel, \cite{ding_enhancing_2024} proposes a biologically inspired defense for SNNs, using stochastic gating to suppress spike patterns introduced by adversarial manipulation or noise. By introducing controlled variability into the spiking activity, the model filters out unusual sequences and improves robustness in the evaluated settings.

\subsubsection{Customized Hardware Design}
Defensive strategies at the hardware level are essential for neuromorphic systems, particularly given the way software-level behavior can be affected by physical leakage, injected faults, and memory-level manipulation. Customized hardware designs aim to embed resilience directly into the architecture, enabling early detection and prevention of attacks while maintaining low overhead in performance and energy. Memristor-based neuromorphic computing, although efficient, is especially vulnerable due to its non-volatile memory properties. Encrypting weights before storage offers a baseline defense, but repeated encryption and decryption can degrade device longevity. Furthermore, common obfuscation methods, such as masking crossbar connections, often incur significant area and power costs, making them impractical for low-resource applications.

To address these challenges, Zou et al. \cite{zou_enhancing_2022} introduce a secure weight-mapping method that selectively encodes weight columns in memristor crossbars using 1’s complement transformations. This encoding obfuscates the true weight distribution unless the mapping key is known, even when conductance values are exposed. The method includes voltage-current–based encoding schemes, on-chip decoding modules, and complementary techniques like decoy weights and crossbar segmentation. Together, these mechanisms provide improved protection against weight extraction while maintaining inference accuracy and system performance, making them a relevant direction for securing memristive neuromorphic devices against model-extraction attacks.


Wong et al. \cite{wong_snngx_2024} propose SNNGX, a novel encryption method for protecting the intellectual property of spiking neural network models implemented on RRAM-based neuromorphic computation. Employing an XOR-based encryption scheme guided by genetic algorithms, SNNGX helps prevent unauthorized access and model cloning, particularly from off-chip memory and side-channel attacks. The method offers low hardware overhead and energy efficiency by avoiding the need to write back decrypted weights to memory. Experimental results confirm its effectiveness, demonstrating strong protection with minimal impact on computational latency and throughput. This makes SNNGX a relevant protection mechanism for deployments where model confidentiality is a primary requirement.

While software-level protections such as encryption and access control are crucial for protecting spiking neural network models, they represent only one part of a comprehensive security strategy. Ensuring the adversarial robustness of SNNs is equally important \cite{ding_snnrat_2022}, and this involves preserving model behavior even when faced with unexpected inputs or adversarial perturbations. Robustness addresses vulnerabilities that can arise during deployment and runtime, complementing techniques that protect intellectual property and data confidentiality \cite{liang_toward_2022}.

Overall, software-level neuromorphic security remains dominated by benchmark-based evaluations on software SNNs using a limited set of static and event-based datasets. Adversarial attacks are the most extensively studied category, whereas temporal backdoors, membership inference, and model inversion have received fewer independent evaluations. Cross-layer side-channel and fault attacks further demonstrate that physical leakage and hardware perturbations can affect model-level behavior, but only a limited number of studies validate these effects on FPGA or dedicated neuromorphic platforms. Existing defenses improve adversarial robustness or protect model intellectual property, yet their costs in clean accuracy, training effort, query complexity, latency, and spike activity are not reported consistently. Future work should therefore establish standardized threat models and evaluation benchmarks that jointly measure attack effectiveness, clean-task performance, computational cost, and transferability to deployed neuromorphic systems. Table~\ref{tab:study-Software} summarizes representative software-level attacks and defenses, together with their evaluation settings and main practical limitations.

\begin{table*}[t]
  \caption{Representative software-level attacks and defenses targeting
  SNNs. SW denotes software-based evaluation, and HW denotes hardware-oriented
  evaluation.}
  \label{tab:study-Software}
  \centering
  \scriptsize
  \setlength{\tabcolsep}{2.5pt}
  \renewcommand{\arraystretch}{1.08}

  \begin{tabular}{@{}
    >{\raggedright\arraybackslash}p{0.60cm}
    >{\raggedright\arraybackslash}p{1.25cm}
    >{\raggedright\arraybackslash}p{2.10cm}
    >{\raggedright\arraybackslash}p{1.30cm}
    >{\raggedright\arraybackslash}p{2.65cm}
    >{\raggedright\arraybackslash}p{2.05cm}
    >{\raggedright\arraybackslash}p{2.45cm}
  @{}}
    \toprule
    \textbf{Paper} &
    \textbf{Class} &
    \textbf{Mechanism} &
    \textbf{Target} &
    \textbf{Dataset} &
    \textbf{Evidence} &
    \textbf{Cost / Limitation} \\
    \midrule

    \multicolumn{7}{@{}l}{\underline{\textit{Adversarial attacks and defenses}}} \\
    \addlinespace[1pt]

    \cite{liang2021exploring}
    & Adversarial
    & Spatio-temporal gradient attack
    & BPTT-trained SNNs
    & MNIST, CIFAR-10, N-MNIST, CIFAR10-DVS, Gesture-DVS
    & SW benchmark
    & White-box; no hardware validation \\

    \cite{bu2023rate}
    & Adversarial
    & Rate-gradient approximation
    & Deep SNN
    & CIFAR-10, CIFAR-100, CIFAR10-DVS
    & SW benchmark
    & White-box; rate-coding specific \\

    \cite{krithivasan_efficiency_2022}
    & Adversarial
    & Accuracy-preserving, spike-activity
    & SNN accelerators
    & CIFAR-10, ImageNet
    & SNN evaluation; 1.7--2.5$\times$ more spikes
    & 1.6--2.3$\times$ latency and 1.4--2.2$\times$ energy; white-box \\

    \cite{yao2024exploring}
    & Adversarial
    & Softmax perturbation of raw event streams
    & SNN vision models
    & CIFAR10-DVS, DVS Gesture, N-MNIST
    & SW benchmarks; 53--87\% targeted ASR
    & White-box; no sensor or hardware validation \\

    \cite{ding_snnrat_2022}
    & Defense
    & Regularized adversarial training
    & Deep SNN
    & CIFAR-10, CIFAR-100
    & SW benchmarks; white- and black-box robustness
    & Additional training cost; BPTT $\approx3\times$ BPTR/CBA \\

    \addlinespace[2pt]
    \multicolumn{7}{@{}l}{\textit{\textit{Backdoor and poisoning attacks}}} \\
    \addlinespace[1pt]

    \cite{abad_sneaky_2024}
    & Backdoor
    & Static and dynamic neuromorphic triggers
    & SNN
    & CIFAR10-DVS, DVS128 Gesture, N-MNIST
    & SW benchmark; up to 100\% ASR with low clean-accuracy impact
    & Trigger-specific \\

    \cite{fu_spikewhisper_2024}
    & Backdoor
    & Distributed temporal triggers
    & Federated SNN
    & N-MNIST, CIFAR10-DVS
    & SW federated benchmark
    & Assumes malicious federated participants \\

    \cite{yadav2024unveiling}
    & Poisoning
    & Malicious training samples
    & SNN
    & MNIST, Fashion-MNIST
    & SW benchmark
    & Limited dataset and deployment diversity \\

    \addlinespace[2pt]
    \multicolumn{7}{@{}l}{\underline{\textit{Privacy attacks}}} \\
    \addlinespace[1pt]

    \cite{moshruba2024neuromorphic}
    & Membership
    & Output-based membership inference
    & SNN/ANN
    & MNIST, CIFAR-10/100, ImageNet
    & SW benchmark
    & Black-box; results depend on training setting \\

    \cite{poursiami2024brainleaks}
    & Inversion
    & Surrogate-gradient spike reconstruction
    & SNN
    & MNIST, AT\&T, N-MNIST, DVS128 Gesture
    & SW benchmark
    & No deployed-hardware validation \\

    \addlinespace[2pt]
    \multicolumn{7}{@{}l}{\underline{\textit{Cross-layer side-channel and fault attacks}}} \\
    \addlinespace[1pt]

    \cite{goswami2024investigation}
    & SCA
    & Differential power analysis
    & FPGA-SNN
    & Fashion-MNIST
    & FPGA power measurements
    & Requires trace acquisition and selected inputs \\

    \cite{nagarajan_fault_2022}
    & Fault injection
    & Voltage/laser-induced neuron faults
    & SNN
    & Digit recognition
    & HW-oriented evaluation
    & Physical feasibility and defense cost limited \\

    \addlinespace[2pt]
    \multicolumn{7}{@{}l}{\underline{\textit{Model and IP protection}}} \\
    \addlinespace[1pt]

    \cite{poursiami_watermarking_2024}
    & IP defense
    & Fingerprint- and backdoor-based watermarking
    & SNN
    & MNIST
    & SW benchmark
    & Sensitive to fine-tuning; limited task diversity \\

    \bottomrule
  \end{tabular}
\end{table*}

\section{Security Enablers in Neuromorphic Devices}

\begin{figure}[ht]
  \centering
  \includegraphics[width=0.8\linewidth]{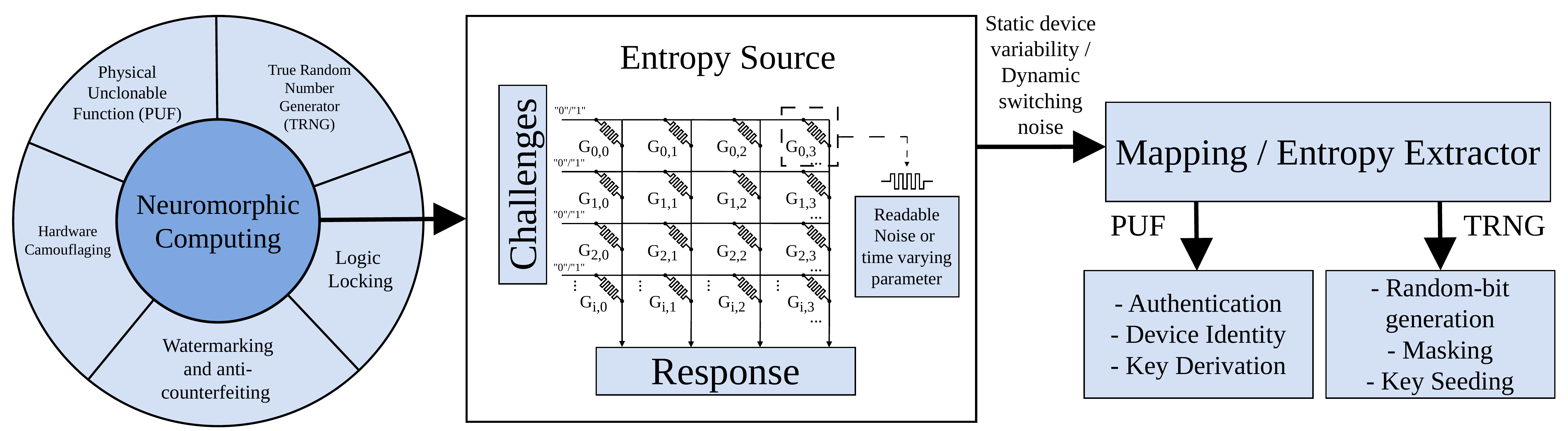}
  \caption{Conceptual use of memristive arrays as security enablers. Static device-to-device variability can provide device-specific responses for PUFs, whereas dynamic switching noise and temporal variability can provide entropy for TRNGs. The extracted responses can support authentication, key generation, and related hardware-security functions.}
  \label{fig:study-PUF}
  \Description{Conceptual diagram showing a memristor array receiving challenges and producing device-dependent responses. Static variability is mapped to Physical Unclonable Function responses, while dynamic switching noise is processed as entropy for True Random Number Generation. The resulting bitstreams support authentication, cryptography, and secret-key generation.}
\end{figure}

Neuromorphic systems combine in-memory computation with device-level plasticity, enabling efficient and adaptive processing. As mentioned in Section~\ref{subsec:inmemory}, \textit{memristors} play a central role in neuromorphic circuit design: they can emulate synaptic learning rules in SNNs, support analog in-memory computing in neuromorphic processors, and provide physical entropy for security mechanisms through their inherent stochasticity and device-to-device variability. Consequently, variability-based primitives such as PUFs and TRNGs are increasingly explored for hardware-level protection, key generation, and identity verification \cite{indiveri_memory_2015, mehonic_brain-inspired_2022_, ham_neuromorphic_2021}. Memristive arrays can serve as device-specific entropy sources, supporting compact security primitives that may be integrated close to memory and computation \cite{florian_using_2022}. These mechanisms exploit static process variation or dynamic switching noise to generate device fingerprints and random bitstreams, although their practical integration depends on readout circuitry, calibration, environmental stability, and error-correction requirements, as illustrated in Fig.~\ref{fig:study-PUF}.

\subsection{Physical Unclonable Functions}

Neuromorphic systems are not only bio-inspired in functionality but also leverage intrinsic device variability to implement security primitives directly within neuromorphic circuits. Among these, PUFs have gained traction for enabling lightweight, hardware-level authentication, device identification, and key generation. In particular, the temporal dynamics of resistive switching and threshold behavior in emerging-memory devices can provide device-specific responses for authentication and key generation. Some recent designs additionally evaluate their resistance to machine-learning modeling and side-channel attacks \cite{ibrahim_resilience_2024}. Neuromorphic components such as leaky integrate-and-fire neurons have also been proposed as PUF building blocks by exploiting device-specific spiking responses \cite{nasab2025late}. However, current evidence for these neuron-based designs is primarily circuit- and Monte-Carlo-simulation based, and their reliability and cost on fabricated neuromorphic platforms remain to be established.

Memristive PUF architectures exploit device-level variability and switching non-linearity to obtain compact challenge--response mechanisms, although their energy, readout, and reliability costs depend on the specific implementation. Shao et al. \cite{shao_imce_2024} report improved resistance to machine-learning modeling attacks by combining non-linear XOR operations with randomized challenge-response pairs (CRPs), achieving near-ideal unpredictability. Similarly, Florian et al. \cite{florian_using_2022} investigate frequency-domain responses of memristive devices as PUF signatures, using CNN-based classification to distinguish device-specific behavior. Their results support the feasibility of extracting distinguishable and repeatable fingerprints from memristive dynamics, although the security relevance depends on reliability, entropy, and resistance to modeling attacks. These studies demonstrate several ways of deriving authentication responses from emerging-memory behavior, but their security must be assessed jointly in terms of uniqueness, reliability, modeling resistance, environmental stability, and implementation cost.
 
In another study, John et al. \cite{john_halide_2021} show that memristors can produce reconfigurable responses due to inherent write-to-write variability, making them suitable entropy sources for secure key generation and device authentication. They further propose a reset-based mechanism that enables response reconfiguration, supporting credential renewal or ownership transfer while reducing dependence on persistent helper data and the need for error-correcting codes. However, long-term reliability, endurance, and environmental sensitivity remain relevant deployment considerations.

By leveraging the intrinsic stochastic properties of memristive arrays, PUF-based designs can support authentication, device identification, and key generation within neuromorphic and emerging-memory hardware. Current evidence shows that these designs are promising, but their maturity remains uneven: many works rely on circuit, Monte Carlo, or device-level evaluations, while fewer provide fabricated-chip measurements under temperature, aging, and repeated-use conditions. Reliability under environmental variation, resistance to modeling and side-channel attacks, and the cost of helper data and error correction also remain inconsistently evaluated. In the following section, we shift focus to TRNG, another security primitive that relies on device-level stochasticity but serves distinct cryptographic roles, particularly in entropy generation, key seeding, and masking.

\subsection{True Random Number Generators and Integrated Security Features}
Another application of the stochastic switching behavior of memristive devices, which results from natural variations in device resistance during write or erase operations, is true random number generation, a mechanism widely used in cryptographic and secure hardware applications \cite{tehranipoor2021emerging, abunahla_memristor_2018, wang2025monolithic}. These approaches obtain entropy from device behavior and may reduce reliance on separate digital entropy sources. Their total cost, however, also depends on sensing, readout, calibration, post-processing, and error-control circuitry \cite{tehranipoor2021emerging}. TRNG quality is commonly evaluated through entropy estimates, output bias, autocorrelation, throughput, and statistical test suites such as NIST SP800-22 and SP800-90B. In contrast, PUFs are typically characterized through uniqueness, reliability, uniformity, bit aliasing, and resistance to modeling attacks \cite{tehranipoor2021emerging}.

To improve entropy generation in secure hardware systems, recent works explore architectures that integrate randomness directly into logic and memory elements. Bende et al. \cite{bende2026rram} demonstrate a CiM-integrated TRNG based on 1T1R RRAM arrays, exploiting cycle-to-cycle variability and read-noise variations as entropy sources. Their design validates randomness using NIST SP800-22 and SP800-90B tests, reaching close to 270 Mbps throughput with an energy cost of 51-66 pJ/bit while exposing a trade-off between readout complexity, endurance, and entropy quality. Wang et al. \cite{wang2025monolithic} extend this direction by presenting a monolithic RRAM-based compute-in-memory array that integrates TRNG, PUF, encryption, and analog computation within the same reconfigurable hardware fabric. This integration can reduce duplicated storage, readout, and control circuitry compared with implementing each function as a separate block, although the resulting area and energy benefits depend on the array organization and peripheral circuits. Ding et al. \cite{ding2021unified} further show that TRNG and PUF functions can be implemented within the same threshold-switching array by exploiting dynamic switching variations for randomness and static leakage-current mismatch for device uniqueness. Their TRNG achieves sub-pJ/bit operation and passes NIST randomness tests without post-processing, while the PUF shows low native bit-error rate, near-ideal Hamming-distance behavior, and resistance to machine-learning attacks. These TRNG and PUF designs show that device-level randomness can support compact security primitives, although practical deployment still depends on entropy stability, environmental robustness, and validation under realistic operating conditions.

Chien et al. \cite{chien_attack_2023} present a TRNG that uses charge trapping dynamics and the stochastic switching behavior of ferroelectric devices to mitigate modeling and signal-analysis attacks. The generator produces unpredictable bitstreams across multiple erase–program cycles, validated by NIST tests and balanced output distributions. Similarly, Dodda et al. \cite{dodda_all--one_2022} propose an integrated sensing and cryptographic engine that uses device-level variability and collective signal processing to encode input data for ultra-low-power encryption. The reported evaluation indicates resistance to decryption under the considered eavesdropper and model-knowledge assumptions, although broader cryptanalytic and implementation-cost evaluations remain necessary.

Overall, emerging-memory devices provide several mechanisms for integrating authentication, entropy generation, and cryptographic support close to sensing, memory, or computation. However, the maturity of the evidence varies substantially, ranging from analytical and Monte Carlo studies to device measurements and a smaller number of fabricated-chip demonstrations. PUF evaluations should jointly consider uniqueness, reliability, environmental stability, modeling resistance, and helper-data or error-correction costs, while TRNG evaluations should report entropy, bias, throughput, energy per bit, aging, and resistance to manipulation or side-channel observation. Statistical randomness tests alone are therefore insufficient to establish practical security. Future work should prioritize standardized evaluation across temperature, voltage, endurance, aging, and repeated-enrollment conditions, together with complete reporting of sensing, readout, post-processing, area, and energy overheads.

Table~\ref{tab:security-primitives-summary} summarizes representative security enablers selected to cover different physical mechanisms, validation settings, security roles, and implementation-cost profiles. A complete paper-by-paper comparison is maintained in the accompanying repository.

\begin{table*}[t]
\caption{Representative security enablers based on neuromorphic and
emerging-memory devices. Direct denotes demonstrated integration with
neuromorphic, compute-in-memory, or in-sensor processing; Compatible
denotes technological relevance without demonstrated integration.
N/R denotes not reported.}
\label{tab:security-primitives-summary}
\centering
\scriptsize
\setlength{\tabcolsep}{2.2pt}
\renewcommand{\arraystretch}{1.06}

\begin{tabular}{@{}
  >{\raggedright\arraybackslash}p{0.75cm}
  >{\raggedright\arraybackslash}p{1.55cm}
  >{\raggedright\arraybackslash}p{2.15cm}
  >{\raggedright\arraybackslash}p{1.45cm}
  >{\raggedright\arraybackslash}p{1.25cm}
  >{\raggedright\arraybackslash}p{2.00cm}
  >{\raggedright\arraybackslash}p{3.75cm}
@{}}
\toprule
\textbf{Paper} &
\textbf{Primitive} &
\textbf{Mechanism} &
\textbf{Technology} &
\textbf{Neuro./CiM} &
\textbf{Evidence} &
\textbf{Metrics / Cost} \\
\midrule

\multicolumn{7}{@{}l}{\underline{\textit{Physical Unclonable Functions}}} \\
\addlinespace[1pt]
    
    \cite{florian_using_2022}
    & PUF
    & Frequency spectra and hysteresis-loop fingerprints
    & Memristor array
    & Compatible
    & Two 16-cell arrays; DFT and CNN classification
    & 97\% accuracy/F1; unstable cells excluded \\
    
    \cite{shao_imce_2024}
    & PUF/CIM encryption
    & Hamming-distance CRPs and masked-key in-situ decryption
    & 2T-FeFET
    & Direct
    & HSPICE circuit/system and ML evaluation
    & 50\% ML accuracy; 289\,fJ/bit and 48$F^2$/cell \\
    
    \cite{ibrahim_resilience_2024}
    & PUF
    & C2C threshold-voltage variability
    & Cu/HfO memristor
    & Compatible
    & Measured-device data; NIST and nine ML attacks
    & All NIST tests passed; 49--52\% ML accuracy \\

    \cite{john_halide_2021}
    & PUF
    & Write-back, C2C reconfiguration, and recurrence
    & Flexible halide-perovskite memristor
    & Direct
    & 1-kb fabricated array; environmental, NIST, and ML tests
    & 0\% BER after write-back; 52\% max. ML accuracy \\
    
    \cite{lin2025re}
    & PUF
    & 3T2R voltage division and inverter-voltage reconfiguration
    & 180-nm ReRAM
    & Compatible
    & Fabricated chip; temperature, DNN, and SCA tests
    & 1\% BER at 85$^\circ$C; $\sim$50\% DNN and $<$70\% SCA accuracy \\
    
    \cite{nasab2025late}
    & Neuron-PUF
    & MTJ process variation in a unified LIF/PUF cell
    & STT-MTJ and 65-nm CMOS
    & Direct
    & Spectre Monte Carlo; 200$\times$128 PUF instances
    & No auxiliary PUF circuit; 50.07\% uniqueness and 0.9974 entropy \\
    
    \addlinespace[2pt]
    \multicolumn{7}{@{}l}{\underline{\textit{True Random Number Generators}}} \\
    \addlinespace[1pt]
    
    \cite{chien_attack_2023}
    & TRNG
    & Ferroelectric switching, charge trapping, and self-correction
    & MoS$_2$/HZO FeFET
    & Compatible
    & Fabricated devices and 8$\times$1 array; NIST and ML tests
    & Entropy 0.99; HD 0.50; ML prediction $\approx$50\% \\
    
    \cite{bende2026rram}
    & TRNG
    & C2C resistance variability, read noise, and shift-XOR
    & 1T1R RRAM-CIM
    & Direct
    & Fabricated 31$\times$16 array; measured currents and emulated readout
    & $\sim$270\,Mbps and 51--66\,pJ/bit; TIA/ADC modeled \\
    
    \addlinespace[2pt]
    \multicolumn{7}{@{}l}{\underline{\textit{Integrated and multifunctional security primitives}}} \\
    \addlinespace[1pt]  
    
    \cite{ding2021unified}
    & PUF/TRNG
    & Static leakage mismatch and dynamic threshold switching
    & Threshold-switching memristive array
    & Compatible
    & Fabricated 8$\times$32 array; NIST, power-noise, and ML tests
    & 0.75\,pJ/bit; 2$F^2$/bit PUF; ML accuracy $\approx$50\% \\
    
    \cite{dodda_all--one_2022}
    & Crypto/Sensing
    & Gaussian-noise injection, spike encoding, and collective decoding
    & Memtransistor
    & Direct
    & Fabricated 8$\times$8 array with 320 devices; DNN eavesdropper tests
    & Hundreds of pJ/engine \\
    
    \cite{wang2025monolithic}
    & PUF/TRNG
    & RRAM write/read noise, bitwise logic, and analog VMM
    & 180-nm 1T1R RRAM-CIM
    & Direct
    & Fabricated 32-kb chip; PUF, TRNG, cipher, and CIFAR-100 tests
    & 357\,Mbps; 69.1\% less area; Energy/latency reduced 92.0/25.9\% \\

\bottomrule
\end{tabular}
\end{table*}

\section{Conclusions and Open Challenges}
Although neuromorphic security is gaining traction, the field remains in an early stage. Unlike conventional systems, neuromorphic security still lacks widely adopted threat models, evaluation protocols, and benchmark suites that connect device-level behavior with SNN-level security outcomes. This survey organizes the literature through a cross-layer structure derived from published evidence, covering hardware threats such as side-channel leakage (power, electromagnetic, timing, and thermal signals), fault injection, and hardware Trojans, as well as software-level attacks such as adversarial perturbations, backdoors, membership inference, and model inversion. However, further work is needed to evaluate how these threat categories interact across implementation layers and validation settings. Future efforts should align threat modeling with device physics, circuit behavior, IMC architectures, SNN dynamics, and deployment requirements, enabling developers to tailor defenses to concrete risk profiles. A key gap is a cross-layer threat model that consistently links device behavior, circuit-level leakage, memory-centric architectures, SNN computation, and application-level security objectives. Application-specific constraints, including energy and latency budgets and fail-safe behavior, are reported inconsistently and therefore remain an open evaluation gap.

Future research in neuromorphic security must analyze system vulnerabilities across attack types, affected components, and implementation settings, with a focus on systematic leakage characterization across heterogeneous device platforms. Comparative studies using standardized benchmarks can clarify which vulnerabilities are fundamental to spiking activity and which arise from device physics. Moreover, multi-device systems that combine different memory technologies within the same accelerator may introduce cross-technology leakage paths not yet fully explored. Although the literature is starting to explore side-channel and fault attacks on SNNs and in-memory computing systems, most of these remain simulation-based or conceptual. There is a gap in translating these attacks to neuromorphic platforms such as Intel Loihi, IBM TrueNorth, or memristor-based crossbars. Advancing neuromorphic security toward practical resilience requires experimental validation of attacks under realistic hardware conditions.

There is currently no standardized benchmark suite for evaluating security in neuromorphic systems. A key direction is the development of open-source benchmarking platforms combining device-level variability with neural simulation, enabling reproducible evaluations of leakage, fault injection, and adversarial robustness within the spiking domain. In addition, the field needs standardized evaluation metrics tailored to each security objective. TRNGs should be assessed using entropy estimates, output bias, autocorrelation, throughput, and statistical test suites such as NIST SP800-22 and SP800-90B, whereas PUFs should be evaluated through uniqueness, reliability, uniformity, bit aliasing, environmental stability, and resistance to modeling attacks. These systems must account for real-world sources of variability, such as thermal drift, radiation impact, and aging effects (all critical to security), to better simulate deployment conditions. Finally, public repositories of tested countermeasures and archived attack logs would support open collaboration and accelerate progress across the community.

Furthermore, implementing energy-aware security protocols that dynamically scale protections based on the available energy budget could be crucial for energy-constrained and latency-sensitive neuromorphic deployments. Trade-offs must also be carefully managed between the area and power overhead introduced by shielding or power balancing techniques and the actual leakage reduction they offer. Another promising direction is to use device-specific PUF responses to authenticate devices and protect weight provisioning during initialization, embedding unique hardware fingerprints into each neuromorphic device. Techniques such as anomaly detection in spike statistics, hardware fuzzing to stress-test rare fault and leakage behaviors, or self-checking neurons to verify spike integrity could provide first steps toward resilience, especially against fault injection and hardware Trojan attacks.

From a purely software perspective, existing studies confirm that SNNs are vulnerable to data poisoning and adversarial perturbations, which can significantly degrade accuracy even in lightweight models. However, evaluations should move beyond MNIST-like datasets and purely simulation-based settings toward hardware-aware studies using detection, segmentation, closed-loop control, and event-camera benchmarks. At the software level, adversarial attacks and defenses are the most extensively evaluated, while temporal backdoors and privacy attacks have fewer independent studies and limited hardware validation. Security enablers such as PUFs and TRNGs include several fabricated demonstrations, but their complete readout, calibration, reliability, and energy costs are not reported consistently. Future work should explore hybrid defenses that combine differential privacy, robust training, SNN watermarking, hybrid IP-protection mechanisms, and poisoned-data filtering, while considering the resource constraints of neuromorphic hardware.

Addressing these challenges requires collaboration across disciplines. Materials scientists can model degradation, thermal drift, and stochastic switching; circuit designers can develop efficient, tamper-resistant architectures; cryptographers can design lightweight PUF and TRNG primitives, and software experts can build SNNs resilient to attacks. Only through such convergence can we achieve secure, scalable, and trustworthy neuromorphic platforms.


\bibliographystyle{ACM-Reference-Format}
\bibliography{references}

\appendix

\end{document}